# Robust organic radical molecular junctions using acetylene terminated groups for C-Au bond formation

Francesc Bejarano,[a] Ignacio Jose Olavarria-Contreras,[b] Andrea Droghetti,[c*] Ivan Rungger,[d] Alexander Rudnev,[e,f] Diego Gutiérrez,[a] Marta Mas-Torrent,[a] Jaume Veciana,[a] Herre S. J. van der Zant,[b] Concepció Rovira,[a] Enrique Burzurí,[b,g*] Núria Crivillers[a*]

[a] Department of Molecular Nanoscience and Organic Materials, Institut de Ciència de Materials de Barcelona (ICMAB-CSIC) and CIBER-BBN, Campus de la Universitat Autonoma Barcelona (UAB), 08193 Bellaterra, Spain

[b] Kavli Institute of Nanoscience, Delft University of Technology, Lorentzweg 1, Delft 2628 CJ, The Netherlands

[c] Nano-Bio Spectroscopy Group and European Theoretical Spectroscopy Facility (ETSF), Universidad del País Vasco (UPV/EHU) , Avenida Tolosa 72, 20018 San Sebastian, Spain

[d] National Physical Laboratory, Teddington, TW11 0LW, United Kingdom

[e] University of Bern, Department of Chemistry and Biochemistry, Freiestrasse 3, 3012 Bern, Switzerland

[f] Russian academy of sciences A. N. Frumkin Institute of Physical Chemistry and Electrochemistry RAS, Leninskii pr. 31, Moscow, 119991, Russia

[g] IMDEA Nanoscience, Ciudad Universitaria de Cantoblanco. c/Faraday 9, 28049 Madrid, Spain

**ABSTRACT:** Organic paramagnetic and electroactive molecules are attracting interest as core components of molecular electronic and spintronic devices. Currently, further progress is hindered by the modest stability and reproducibility of the molecule/electrode contact. We report the synthesis of a persistent organic radical bearing one and two terminal alkyne groups to form Au-C σ bonds. The formation and stability of self-assembled monolayers and the electron transport through single-molecule junctions at room temperature have been studied. The combined analysis of both systems demonstrates that this linker forms a robust covalent bond with gold and a better-defined contact when compared to traditional sulfur-based linkers. Density functional theory and quantum transport calculations support the experimental observation highlighting a reduced variability of conductance values for the C-Au based junction. Our findings advance the quest for robustness and reproducibility of devices based on electroactive molecules.

## 1. INTRODUCTION

The molecule/electrode contact plays a fundamental role in the performance of molecular electronic devices since it directly affects the charge transport across the interface.[1–3] The search for a more stable molecule-electrode bond, a well-defined interface geometry and more conductive interfaces is the driving force to pursue robust and efficient molecule based devices.[4] The chemisorption of molecules on noble-metal electrodes has been mainly achieved through thiols,[5] but also, by employing other groups such as pyridines,[6–9] amines,[10–13] N-heterocylic carbenes,[14] isothiocyanide[15] and carbodithioate.[16] Interestingly, some recent works have shown that the formation of covalent highly directional σ-bonded C-Au junctions provides high conductance at the single-molecule level. Different anchoring groups such as trimethyl tin ($SnMe_3$)-terminated polymethylene chains[17] and silyl-protected acetylenes[18–20] have been employed for this aim. In the case of the silyl groups there is the need of applying in situ desilylation chemistry to form the Au-C bond. On the contrary, the terminal acetylene (R-C≡C-H) group spontaneously forms stable C-Au bonds.[21,22] This strategy has been mainly used to prepare self-assembled monolayers (SAMs) on Au (flat surfaces[23–26] and Au nanoparticles[27]) and on Ag.[23,24] Charge transport measurements through some of these SAMs have been performed in large-area molecular junctions (using $Ga_2O_3/E_{GaIn}$ as top electrode), in an STM break-junction,[26] and with electrochemical scanning tunneling spectroscopy.[28] Recently, a mechanically controlled break-junction (MCBJ) technique was used to compare the single-molecule conductance of a family of alkynyl terminated oligophenylenes.[29] However, up to now, single-molecule measurements through bridges incorporating a functional moiety linked via a -C≡C-Au bond at room temperature (RT) are very scarce.

In the present work, we therefore aim at filling this fundamental knowledge gap. To this objective, we exploit the highly persistent perchlorotriphenylmethyl (PTM) radical as a functional moiety. We show that the -C≡C-Au bond allows for drastic improvements in the reproducibility of conductance measurements and bond strength when compared to other commonly used contacts. We demonstrate that the magnetic character is preserved after covalent bonding. The charge transport mechanism of PTM radicals in the tunneling regime was previously addressed in SAMs covalently grafted to Au through a thiol group showing that the single-unoccupied molecular orbital (SUMO) was playing a crucial role in the transport enhancing the junction conductivity.[30–32] Electrochemical gating was also used to achieve a highly effective redox-mediated current enhancement.[33] Being all-organic, PTM radicals present an intrinsic magnetic moment, low spin-orbit coupling and low hyperfine interactions. These attractive redox and magnetic properties, absent in transition metal-based magnetic compounds, have recently attracted attention in molecular (spin)electronics[34–36] where long spin coherence times are required to preserve the information encoded in the electronic spin. The individual spin of different radical species has been detected in low-temperature electron transport measurements in the shape of Kondo correlations in molecular junctions[37,38] and surfaces.[39,40] The PTM radical has proved to be robust in the junction thanks to the encapsulation of the radical spin in a three chlorinated-phenyl shell. Moreover, it has been proved that the magnetic state of PTM-based polyradicals can be mechanically modified[41] and electrically gated to form the basis of a quantum SWAP gate.[42] Similar oligo(p-phenyleneethynylene) (OPE)-based radicals have shown large magneto-resistance effects that could be used to tune charge transport in metal-molecule junctions.[43] Moreover, although it remains still to be proved, organic radicals could act as spin filters[44] provided that a strong hybridization between spin and electrodes is achieved. Aside from the magnetic properties, the redox properties of similar organic radicals have been used to enhance charge transport in molecular junctions.[45] All these intriguing properties depend up to some extent on the reproducibility and strength of the bond radical to the electrodes, a key-step that we improve here.

Two novel PTM radical derivatives bearing one (**1-Rad**) and two (**2-Rad**) acetylene terminated groups (Figure 1a) have been designed and synthesized in order to form SAMs and single molecule junctions, respectively. Thanks to the stability of the alkynyl group, compared to the –SH one, there is no need of protecting and deprotecting it during the deprotonation and oxidation reactions required to generate the radical species. Additionally, these linkers once formed are stable without showing signs of oxidation in time, as it is the case of most of the thiolated compounds.[46–48] SAMs based on **1-Rad** (Figure 1b) were successfully prepared. Several characterization techniques, Electron Paramagnetic Resonance spectroscopy (EPR), X-Ray Photoelectron Spectroscopy (XPS), Raman spectroscopy and Cyclic Voltammetry (CV) show the formation of a very stable metal-molecule covalent bond where the unpaired spin is preserved after bonding. We have further downscaled to the single-molecule level. Molecular junctions based on **2-Rad** (Figure 1b) were prepared and compared with the equivalent bisthiophene-terminated derivative (**3-Rad**, Figure 1a) that shares the same functional core but is functionalized with thiophene anchoring groups.[37] Room temperature electron transport measurements and the statistical analysis show that **2-Rad** form a very stable bond with a better defined anchoring geometry when compared to the S-Au bond in the **3-Rad**, while still having similar current levels. These findings are supported by density functional theory (DFT) and quantum transport calculations that predict a C-Au bond three times stronger than the S-Au and with a lower anchoring geometry variability.

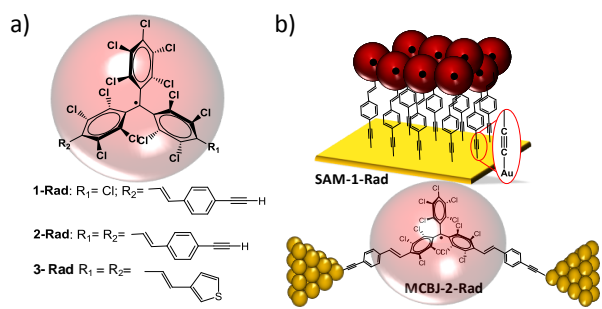

**Figure 1.** a) Chemical structure of the PTM radicals employed in this study. b) Scheme of the SAM based on **1-Rad** (top) and MCBJ based on **2-Rad** (bottom).

## 2. RESULTS AND DISCUSSION

**2.1. Synthesis and Self-Assembled Monolayers Preparation and Characterization.** Molecules **1-Rad** and **2-Rad** were obtained through a two-step reaction from the hydrogenated precursors **1-H** and **2-H** involving the formation of a carbanion on the α carbon and the subsequent one-electron oxidation of this carbanion to its corresponding open-shell (radical) form (Scheme S1 in the Supporting Information (SI)). **1-H** and **2-H** were previously synthesized through a Horner-Wadsworth-Emmons reaction between the monophosphonate-PTM derivative (**3**)[49] and the bisphosphonate-PTM derivative (**4**),[50] respectively, and 4-ethylnylbenzaldehyde (Scheme S1). Closed- and open-shell molecules were fully characterized as described in the SI.

First, self-assembled monolayers (SAMs) of **1-Rad** were prepared on Au(111) to investigate the formation of the C-Au bond. The SAM formation conditions were optimized to maximize the surface coverage (see Supporting Information). SAMs were prepared under inert conditions to avoid the possible oxidation of the alkyne group in presence of $O_2$ and Au, as previously reported.[25]

At this point, it is worth mentioning the reactivity of the terminal acetylenes with the gold substrate. Maity et al. studied the functionalization of gold clusters with a series of terminal alkynes derivatives by means of various spectroscopic methods. They clearly demonstrated the binding motif and the loss of the terminal H through the heterolytic deprotonation of the alkyne as the key mechanism for the binding of the alkynyl group to gold.[22] Furthermore, Raman spectroscopy has been used by several authors to identify the covalent C-Au formation.[18,51,52] Taking all this into account, SAM-**1-Rad** was characterized by Raman spectroscopy (Figure 2a). The spectrum of the **1-Rad** in powder was also acquired for comparison. The weak band corresponding to the C-C triple bond stretching is observed both in powder (2108 $cm^{-1}$) and red-shifted on surface (2035 $cm^{-1}$) confirming the integrity of the alkyne group after the monolayer formation. The displacement of the above-mentioned band is in agreement with previous studies describing the reactivity of terminal alkynes on gold.[22] Remarkable is the appearance of a new sharp band at 432 cm$^{-1}$ in the SAM spectrum. According to literature,[18,51,52] this band can be assigned to the characteristic C-Au stretching mode, and constitutes a strong proof of the covalent character of the binding between the PTM **1-Rad** and the gold surface. These observations point to an up-right configuration of **1-Rad** molecules maintaining the directional triple bond and losing a hydrogen atom. In addition, X-ray photoelectron spectra (XPS) confirmed the presence of chlorine (Figure 2b; see Figure S3 for the C1s spectrum). The Cl2p spectrum showed the typical doublet corresponding to Cl2p$_{1/2}$ and Cl2p$_{3/2}$ at 202.4 eV and 200.8 eV, respectively.

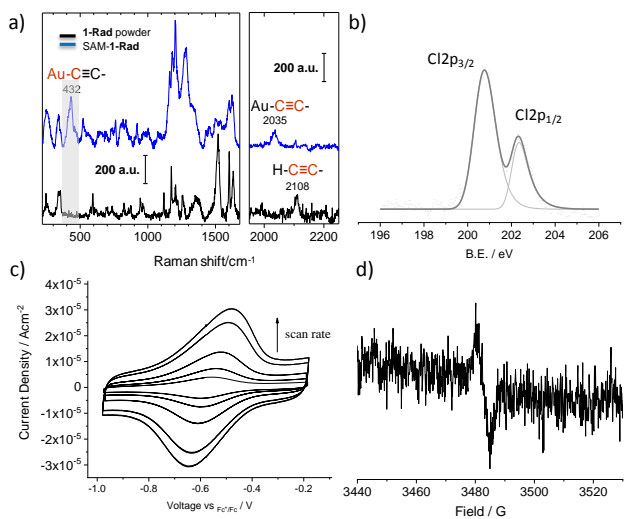

**Figure 2.** Characterization of SAM **1-Rad**. a) Raman spectrum of **1-Rad** in powder (black) and SAM **1-Rad** (blue). b) High resolution Cl2p XPS spectrum. c) Cyclic voltammetry at different scans rates (100, 200, 400, 800, 1000 mV/s) and d) EPR spectrum of a SAM **1-Rad**.

The contact angle value of the SAM **1-Rad** was found to be 79 ± 4°, which is in agreement with previously reported PTM-terminated SAMs.[53,54] Moreover, the multifunctional character of the PTM radical molecules allows using several techniques to confirm the successful grafting and the monolayer robustness. In the cyclic voltammetry a single clearly reversible peak at -0.57 V (vs Fc$^+$/Fc) can be observed for the SAM-**1-Rad** (Figure 2c), confirming the redox-active character of the monolayer. The observed peak's separation ($\Delta E$) is attributed to lateral interactions between the electroactive PTM moieties.[54] As shown in Figure S2a, the current linearly increased with the scan rate, which is indicative of surface confined species. The surface coverage was found to be $8 \pm 1 \cdot 10^{-11}$ mol/cm$^2$ (see SI for further details on the calculation). The stability of the layer under electrochemical conditions was evaluated by performing 20 consecutive cycles while sweeping the bias between -1 V and -0.2 V without showing a decrease of the current intensity in the redox peaks (Figure S2b). Finally, the paramagnetic character of the layer, and therefore the persistence of the unpaired spin in the **1-Rad** after cova-lent anchoring, was proved by EPR spectroscopy (Figure 2d). The signal can be fitted with a *g* value of 2.0024, in agreement with that obtained for PTM radical in solution, and a line width of 4.3 G. Thanks to the combination of these techniques, we clearly demonstrate that the redox-active **1-Rad** molecule covalently linked to the surface preserves the unpaired spin.

In addition to the chemical and structural characterization of the **1-Rad** SAMs, charge transport measurements across the monolayer were performed by top-contacting the layer with a $^E$GaIn based electrode. Similar results to the ones previously reported with thiolated SAMs were obtained[32] (see SI for further details, Figure S4). The high reproducibility of these measurements supports the feasibility of the grafting approach to obtain stable monolayers through the C-Au bond.

**2.2. Single-Molecule Junctions.** We have further downscaled to study the bond and the conductance at the single-molecule level. We have performed RT electron transport measurements through individual **2-Rad** molecules using the (MCBJ) technique described in Refs [55,56] and in the SI. Figure 3a shows a two-dimensional conductance vs electrode displacement histogram for **2-Rad** molecules. The histogram is made using 2500 consecutive traces recorded at an electrode speed of 6 nm/s and *V* = 0.2 V. Two characteristic features are present: first, the conductance drops exponentially from 10$^{-4}$G$_0$ to the noise level for electrode displacements between 0 and 0.7 nm. These traces are typically attributed to empty junctions with no molecules bridging the electrodes. The second feature is a single flat plateau with a characteristic length of 2.3 nm (see length histogram in Figure S5) after which the conductance drops abruptly. Individual examples of these traces can be observed in the inset of Figure 3a. The plateau is indicative of the formation of a molecular junction that breaks when the gap between the electrodes is too wide for the **2-Rad** molecule to bind. A significant 30% of the traces show molecular signatures. Interestingly, the plateau length plus the snap-back correction (~0.5 nm) approximately matches the length of the relaxed **2-Rad** molecule bonded to Au (2.7 nm).

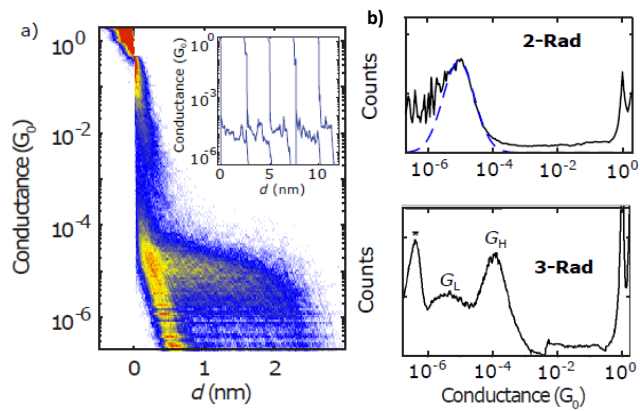

**Figure 3.** (a) Two-dimensional conductance vs electrode displacement histogram of the **2-Rad** molecule constructed from 2500 consecutive traces at RT and 0.2 V bias voltage. The inset shows some selected individual traces. (b) One-

dimensional conductance histogram for **2-Rad** (top) and **3-Rad** (bottom) constructed by integrating over the displacement in (a) and Figure S8b, respectively.

The conductance value of the plateau in Figure 3a is determined from the one-dimensional conductance histogram shown in Figure 3b. The dotted line is a log-normal distribution fit around the conductance regions displaying molecular features. The most probable conductance value for the **2-Rad** molecule is $8.8 \times 10^{-6}$ $G_0$.

To gain a deeper insight, we have compared these results with the conductance characteristics of a similar PTM moiety functionalized with thiophene anchoring groups (**3-Rad**, Figure 1) measured with the Scanning Tunnelling Microscope (STM)-based break-junction (STM-BJ) technique. This molecule was synthesized as previously reported.[37] As seen in Figure 3b (bottom), in the latter case, the conductance values appear more spread between two conductance plateaus at $G_{high} = 10^{-4} G_0$ and $G_{low} = 4 \times 10^{-6} G_0$ (see also Figure S8). The asterisk indicates the noise level of the set-up. The small spike at log(G/G0) ~ -2.2 in panel (b) bottom is an artefact related to the switching of the amplifier stage. This spread of values is signature of a weaker or ill-defined molecule-electrode contact[57] in contrast with the remarkably well defined plateau measured in **2-Rad** junctions, signature of a better-defined molecule-electrode anchoring geometry. Moreover, the size of the plateau (plus snap-back) in the **3-Rad** junction extends only up to 0.95 nm, whereas the length of the fully stretched molecules is 1.7 nm, suggesting the molecules are anchored adopting different electrode/molecule configurations (see length histograms of both species in the SI). These results point to a stronger mechanical bond in the case of **2-Rad** that allows fully stretching of the molecule before breaking the junction and points to a preserved structural integrity of the molecule. Note that the strength in the mechanical bond is not straightforwardly translated into a higher conductance.

Finally, the only sensitive tool to detect the radical (unpaired) spin at the single molecule level is through its interaction with the conduction electrons in the shape of Kondo correlations.[37] These correlations are however only significant at very low temperatures for PTM radicals (<10K).[37] Figure S6 shows some selected dI/dV curves measured at T = 4 K in a MCBJ setup. The current is measured at a fixed inter-electrode distance and the differential conductance is thereafter numerically obtained. A zero-bias resonance is observed in some of the junctions. This resonance is typically associated to Kondo correlations in molecules or quantum dots containing magnetic impurities and appears when the electronic hybridization of impurity and conduction electrons is strong.

**2.3. Theoretical Calculations.** The strength of the mechanical bond between the molecule and Au is quantified by means of the maximum rupture force $F$ and of the bond dissociation energy $D$.[58] These cannot be directly obtained from our experiments, but they can be estimated by calculating the potential energy surface (PES) for the stretched bond with DFT, and then fitting the PES with a Morse potential (details are given in the SI, Figures S11-S13).

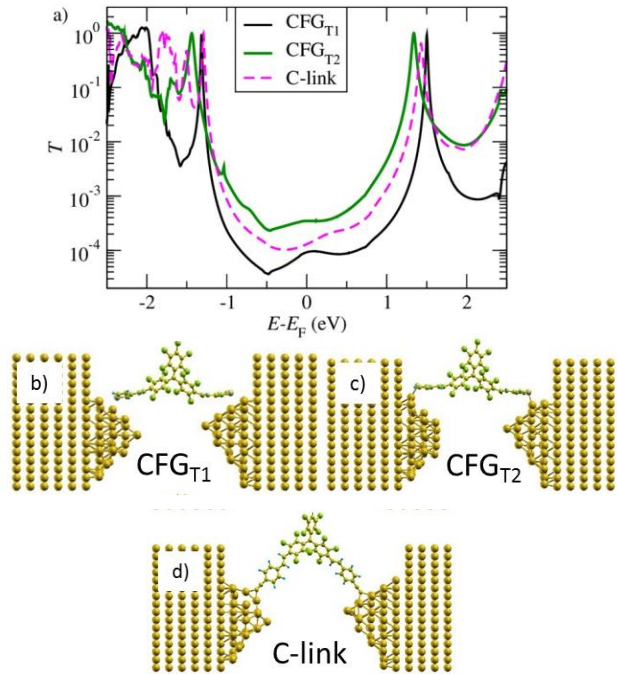

**Figure 4.** (a) Transmission as function of energy for different atomic configurations (CFGs): (b) CFG T1 and (c) CFG T2 are two typical structures expected to be found in high temperature break-junction measurements with **3-Rad** and Au; (d) with **2-Rad** directly bonded via C-Au link.

We find that the C-Au bond forms at a bridge adsorption site, where $F$= 3.05 nN and $D$= 3.5 eV. The energy is about 300 meV lower than for the top adsorption site, while the hollow adsorption site is not stable as the geometry optimization ends up the bridge configuration without going over any potential barrier. Most importantly, such Au-C bond has to be elongated by about 0.5 Å under an applied force as large as 2.9 nN for the Au-C bond-breaking activation energy to equal RT (Figure S11). This means that in RT MCBJs experiments the Au-C bond will not break until it is stretched up to 0.5 Å. Overall the results support the observations about the mechanical junction stability in Figure 3. In contrast, as described in the SI, a thiophene linker establishes no covalent bond with a flat Au surface, so that it is physisorbed (Figure S9-S10). A covalent bond can only be established between the thiophene S atom and an Au adatom on a corrugated surface (Figure S13). In this case, for the adatom-S bond, $F$ and $D$ are just 0.45 nN and 0.3 eV, respectively, and the bond-breaking activation energy equals RT when the bond is stretched by less than 0.1 Å. For comparison, we also point out that for the S-Au bond of standard thiol linkers used in previous experiments (Figure S12),[32] $F$ and $D$ are about three times smaller than for the Au-C bond, i.e. $F$=1.45 nN and $D$=1.26 eV.

Finally, we also studied the conductance of junctions comprising **2-Rad** and **3-Rad** molecules by using DFT-based quantum transport calculations (details are given in SI and results in Figure 4). **3-Rad** can contact either a rather flat part of the electrode or a protruding adatom, and we model these two situations by considering different atomic configurations (CFGs), for instance configuration CFG T1 and CFG T2 in

Figure 4b-c. In the first case, as outlined above the bond to the Au is non-covalent, and the molecule can therefore slide on the Au tip as the junction is elongated. The conductance is quite low due to the weak electronic coupling between Au and the molecule. For more corrugated Au tips, the covalent bond between the thiophene S atom and the protruding Au adatom results in an increased electronic coupling and therefore in a higher conductance. This is in agreement with the experimental results, where for **3-Rad** a quiet wide range of conductances between a higher and a lower value is found (Figure 3b and Figure S8).

For **2-Rad**, whenever the relative displacement of the electrodes is in the plateau region of Figure 3, the linkers establish a strong covalent bond with the electrodes, where the C atom is on the bridge site between two Au atoms (Figure 4d), in an analogous way to what found for the adsorption on flat Au surfaces (see SI). The fact the molecule binds to the electrodes in such a locally well-defined C-Au bond, independently of the overall junction geometry, drastically reduces the variability in conductance values. The calculated transmission curve lies in an intermediate range between the ones for CFG T2 and CFG T1. The low bias conductance corresponds to the transmission at the Fermi energy, for which we obtain about $10^{-4}$. This is on the upper end of the experimentally measured conductance peak (Figure 3).

## 3. CONCLUSIONS

In conclusion, the functionalization of a PTM organic radical with alkynyl end groups has led to the formation of a robust covalent binding between these electroactive and paramagnetic molecules and Au. Throughout a detailed comparison with a similar thiophene functionalized derivative we prove that the Au-C bond provides a more robust and better-defined anchoring geometry as supported by DFT calculations. Our findings open the door to more reproducible spintronics devices based on multifunctional molecules.

## ASSOCIATED CONTENT

**Supporting Information**. Synthesis of the compounds, SAM characterization, electron transport measurements on PTM-bisthiophene and details of the DFT calculations.

## AUTHOR INFORMATION

**Corresponding Author**

\* andrea.droghetti@ehu.eus; enrique.burzuri@imdea.org; ncrivillers@icmab.es

## ACKNOWLEDGMENT

We acknowledge Dr. G. Sauthier from the ICN2 for the XPS measurements, Prof. Carlos Gomez from IcMOL for the SQUID measurements, A. Bernabé and Dr. V. Lloveras from ICMAB for the LDI-ToF and EPR measurements, respectively. This work was supported by FET ACMOL project (GA no. 618082), CIBER-BBN, the DGI (Spain) project FANCY CTQ2016-80030-R, the Generalitat de Catalunya (2014-SGR-17) and the MINECO, through the "Severo Ochoa" Programme for Centers of Excellence in R&D (SEV-2015-0496). F.B he is enrolled in the Materials Science Ph.D. program of UAB. F. B thanks the Ministerio de Educación, Cultura y Deporte for the FPU fellowship. We thank the Dutch science foundation NWO/FOM for financial support.

Supporting Information for

**Robust organic radical molecular junctions using acetylene terminated groups**

**for C-Au bond formation**

Francesc Bejarano, Ignacio Jose Olavarria-Contreras, Andrea Droghetti, Ivan Rungger, Alexander Rudnev, Diego Gutiérrez, Marta Mas-Torrent, Jaume Veciana, Herre S. J. van der Zant, Concepció Rovira, Enrique Burzurí, Núria Crivillers

TABLE OF CONTENTS

# 1. Materials

4-ethynylbenzaldehyde and aqueous 54-56% tetrabutylammonium hydroxide were purchased from Sigma-Aldrich. *p*-chloranil was purchased from Panreac. Potassium *tert*-butoxide was purchased from Merck. All reagents were used without further purification. Diethyl (4-(bis(perchlorophenyl)methyl)-2,3,5,6-tetrachlorobenzyl)-phosphonate [1] and Tetraethyl ((((perchlorophenyl)methylene)bis(2,3,5,6-tetrachloro-4,1-phenylene))-bis(methylene))bis(phosphonate) [2] were synthesized following the reported procedures. All solvents were of HPLC grade and, in the case of toluene, it was distilled over Na/benzophenone prior to its use. Au(111) (300nm) evaporated on mica substrates were purchased from Georg Albert Physical Vapor Deposition.

# 2. Apparatus

## 2.1. Raman spectroscopy

For SERS measurements a polycrystalline gold electrode disk (1.6 mm diameter, BASi) in a solvent-resistant polychlorotrifluoroethylene body was polished with alumina suspension, first 1 μm and then 0.3 μm particles. Then, the electrode was sonicated and thoroughly rinsed with Milli Q water. Afterwards the Au electrode was undergone the electrochemical treatment for roughening the surface. The treatment protocol is based on that reported in [3]. Briefly, we performed 25 potential cycles in 0.1 M KCl (Figure S1) between -0.3 and 1.2 V. Then, the electrode was thoroughly rinsed with Milli Q water and placed into a freshly prepared **1-Rad** solution to obtain SAM (protocol is the same as described in the text). After assembly, the electrode was taken out, sonicated, washed with toluene, dried in an argon stream, and transferred to Raman set-up.

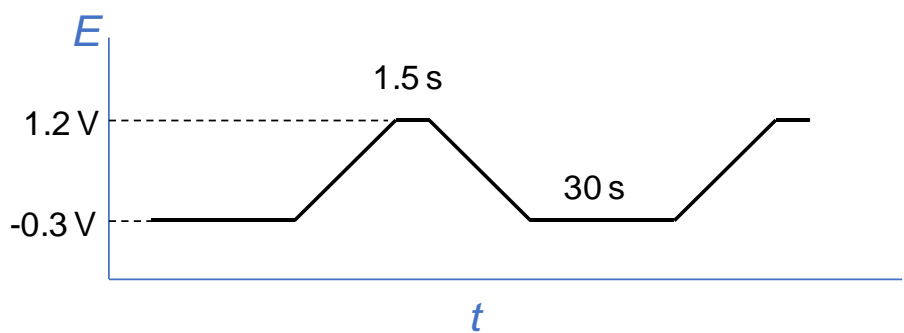

**Figure S1.** Illustration of polarization program for roughening the Au electrode. The shown cycle is repeated 25 times. The ramping rate between -0.3 and 1.2 V was 0.05 V s$^{-1}$.

SER spectra were recorded with a LabRam HR800 confocal microprobe Raman system (HORIBA, Jobon Yvon). The excitation wavelength was 632.8 nm from a He-Ne laser. The power on samples was 0.012 mW. Such a low power was used in order to exclude the decomposition of our compound. A 50× magnification long-working-distance objective (8 mm) was used to focus the laser onto samples and to collect the scattered light in a backscattering geometry. The Raman spectra were recorded in ambient conditions. We collected 8-10 spectra at different areas of **1-Rad**/Au and of **1-Rad** powder samples. The spectra were reproducible, and representative ones are shown in this paper.

## 2.2. X-ray photoelectron spectroscopy (XPS)

XPS measurements were performed with a Phoibos 150 analyzer (SPECS GmbH, Berlin, Germany) in ultra-high vacuum conditions (base pressure 5E-10mbar) with a monochromatic aluminium Kalpha x-ray source (1486.74eV). The energy resolution as measured by the FWHM of the Ag $3d^{5/2}$ peak for a sputtered silver foil was 0.58 eV.

## 2.3. Cyclic Voltammetry (CV)

CV characterization was performed with an AUTOLAB 204 with NOVA 1.9 software. We used a custom built electrochemical cell with a Pt wire as counter electrode, Ag wire as pseudo reference electrode, ferrocene as internal reference and the modified Au(111) on mica as working electrode. The area exposed to the tetrabutylammonium hexafluorophosphate (TBAPF$_6$) in tetrahydrofuran electrolyte solution (0.1 M) was 0.5024 cm$^2$.

## 2.4. Electron Paramagnetic Resonance (EPR)

EPR spectra were registered at room temperature on a Bruker ESP 300 E spectrometer provided with a rectangular cavity T102 working with a X band (9.5 GHz). The signal-to-noise ratio of spectra was increased by accumulation of scans using the F/Flock accessory to guarantee large field reproducibility. Precautions to avoid undesirable spectral distortion and line broadenings, such as those arising from microwave power saturation and magnetic field over modulation, were also taken into account to improve sensitivity. For the spectrum shown in Figure S2d in the main text, 15 scans were acquired. The gold/mica substrate is placed parallel to the field. A blank experiment with non-functionalized gold/mica substrate was performed in the same conditions and showed a negligible signal.

## 2.5. UV-vis absorption spectroscopy

UV-Vis spectra were registered on a Varian Cary 5000 UV-Vis-NIR spectrometer. Quartz cuvettes with an optical path of 1 cm were used

## 2.6. SQUID magnetometry

Variable temperature magnetic susceptibility measurements were carried out in the temperature range 2-300 K in cooling and warming scans with an applied magnetic field of 0.5 T on ground single crystals of compounds 1-2 with a Quantum Design MPMS-XL-5 SQUID magnetometer. The isothermal magnetizations were performed on the same samples at 2 K with magnetic fields up to 5 T. The susceptibility data were corrected for the sample holders previously measured under the same conditions and for the diamagnetic contributions as deduced by using Pascal´s constant tables.[4]

## 2.7. High-Performance Liquid Chromatography (HPLC)

HPLC analyses were carried out in reversed-phase mode using an Agilent 1260 Infinity II chromatograph provided with a diode array detector WR and a quaternary pump VL. The column used was TRACER EXCEL 120 ODSA 5 μm 25x0.46 from Teknokroma. The same non-isocratic method was used for all compounds with an acetonitrile/chloroform mixture (total time = 15 min; t = 0 min 90:10, t = 7 min 60:40, t = 13 min 60:40, t = 14 min 90:10).

## 2.8. LDI-ToF mass spectroscopy

Spectra were registered on a Bruker Ultraflex mass spectrometer by operating at ion pulsed extraction in negative mode at high power.

## 2.9. Nuclear magnetic resonance spectroscopy (NMR)

The $^{1}$H-NMR and $^{13}$C-NMR spectra were registered on a Bruker Avance III 400SB spectrometer and calibrated using residual undeuterated dichloromethane ($\delta(^{1}H)$ = 5.32 ppm; $\delta(^{13}C)$ = 53.84 ppm) and residual undeuterated chloroform ($\delta(^{1}H)$ = 7.26 ppm; $\delta(^{13}C)$ = 77.00 ppm) as internal references. The data analysis was carried out with MestReNova software (MestReLab Research S. L.). The following abbreviations were used to designate multiplicities: br = broad signal, s = singlet, d = doublet, m = multiplet.

*2.10. Infrared spectroscopy (IR)*

Spectra were registered with a FT-IR PerkinElmer spectrometer with a diamond ATR accessory.

*2.11. Contact angle*

All measurements were performed using ultrapure water with a Krüss G10 Contact Angle Measuring Instrument, provided with a CCD camera and with software 1.51 for the drop shape analysis. The drop adjustment was done by using DPA32 Tangent 1 and Tangent 2 methods.

## 3. General procedures

Reactions to obtain **1-H** and **2-H** were monitored by thin-layer chromatography (TLC) carried out on 250 μm Sigma-Aldrich silica gel plates (60F-254) using UV light as visualizing agent and a basic solution and heat as developing agents. Reactions to obtain **1-Rad** and **2-Rad** were monitored by UV-Vis spectroscopy following, for the first step, the apparition of two broad bands around 520 and 580 nm and, for the second step, their vanishing and the apparition of a sharp band around 385 nm. Purification of all compounds was carried out by flash column chromatography using Carlo Erba silica gel (60, particle size 35-70 μm).

## 4. Synthesis details of the PTM derivatives

**Scheme S1. Synthesis of compounds 1 and 2 in their closed-shell (1- and 2-H) and open-shell forms (1- and 2-Rad).**

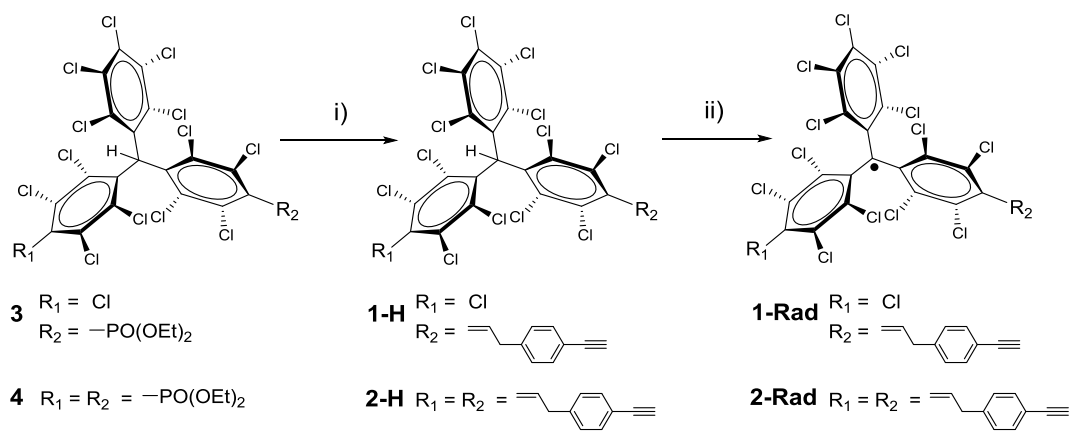

[a]Reagents and conditions: (i) t-BuOK/THF then 4-ethynylbenzaldehyde (-78 °C to rt), (ii) Bu$_4$NOH/THF then p-chloranil (rt)

### 4.1. Synthesis of compound 1-H

Diethyl (4-(bis(perchlorophenyl)methyl)-2,3,5,6-tetrachlorobenzyl)-phosphonate (301 mg, 0.343 mmol, 1 eq) is placed in a previously dried 25 mL Schlenk flask under argon atmosphere. Dry THF (6 mL) is added to dissolve the reagent and the flask is placed in an acetone/liquid N$_2$ bath at -78° C. Afterwards, potassium *tert*-butoxide (64.7 mg, 0.577 mmol, 1.7 eq) is added and the solution turns to yellow. 30 minutes after the addition, when the colour is strongly yellow-orange, 4-ethynylbenzaldehyde (85.6 mg, 0.658 mmol, 1.9 eq) dissolved in dry THF (4.5 mL) is added dropwise and the solution is let to return to room temperature and to stir overnight. The dark solution is quenched upon addition of aqueous 0.4 M HCl drops until a yellow-brown suspension is obtained. Water (8 mL) is added and the product is extracted with CH$_2$Cl$_2$ (3 x 8 mL). The combined organic layers are dried over anhydrous magnesium sulfate, filtered off and concentrated under vacuum. The final crude is separated by column chromatography (SiO$_2$, hexanes), affording the product as a yellowish powder (221 mg, 75% yield). **$^1$H NMR (CD$_2$Cl$_2$, 400 MHz):** δ = 7.53 (br s, 4H, ArC-**H**), 7.11 (d, 1H, $^3J$=16.5 Hz, **H**-C=C-H), 7.06 (d, 1H, $^3J$=16.5 Hz, H-C=C-**H**), 7.04 (s, 1H, **α H**), 3.22 (s, 1H, C≡C-**H**) ppm; **$^{13}$C NMR (CD$_2$Cl$_2$, 101 MHz):** δ = 137.78, 137.31, 136.68, 136.66, 136.50, 136.36, 135.25, 135.18, 135.02, 134.11, 134.08, 133.81, 133.79, 133.69, 133.66, 133.55, 132.74, 132.64, 132.63, 132.43, 126.96, 124.20, 122.67, 83.55 (C≡C-H), 78.68

(C≡C-H), 56.77 (**α C**); **ATR-IR:** ν (cm$^{-1}$) = 3299 (w) (C≡C-H), 3082 (w) (ArC-H), 3032 (w) (C=C-H), 2106 (w) (C≡C), 1635 (w) (C=C), 1605 (w) , 1534 (w) (ArC-ArC), 1504 (w) (ArC-ArC), 1414 (w) (Cl-ArC-ArC-Cl), 1360 (m) (Cl-ArC-ArC-Cl), 1339 (m) (Cl-ArC-ArC-Cl), 1293 (s), 1263 (m), 1240 (w), 1140 (w), 1105 (w), 1017 (w), 967 (w), 947 (w), 872 (w), 859 (w), 847 (w), 812 (s) (ArC-Cl), 807 (w), 765 (w), 739 (w), 713 (w), 699 (w), 688 (w), 671 (m), 649 (m), 614 (w), 610 (w), 578 (w), 539 (w), 530 (w), 519 (w), 506 (w), 502 (w); **LDI-ToF (negative mode):** *m/z* 850.9 [M – H], 779.9 [M – 2 Cl – H], 744.9 [M – 3 Cl – H]; **UV/Vis (THF):** λ (nm) (ε) 241 (49005), 254 (31905), 313 (32129)

## *4.2. Synthesis of compound 2-H*

Tetraethyl ((((perchlorophenyl)methylene)bis(2,3,5,6-tetrachloro-4,1-phenylene))-bis(methylene))bis(phosphonate) (502 mg, 0.506 mmol, 1 eq) is placed in a previously dried 50 mL Schlenk flask. Under argon atmosphere, dry THF (6 mL) is added to dissolve the reagent and the flask is placed in an acetone/liquid N$_2$ bath at -78° C. Afterwards, potassium *tert*-butoxide (140 mg, 1.248 mmol, 2.5 eq) is added and the solution turns to yellow. 20 minutes after the addition, when the colour is strongly yellow-orange, 4-ethynylbenzaldehyde (210 mg, 1.613 mmol, 3.2 eq) dissolved in dry THF (4.5 mL) is added dropwise and the solution is let to return to room temperature and to stir overnight. The dark solution is quenched upon addition of aqueous 0.4 M HCl drops until a yellow-brown suspension is obtained. Water (20 mL) is added, and the product is extracted with CH$_2$Cl$_2$ (3 x 15 mL). The combined organic layers are dried over anhydrous sodium sulfate, filtered off and concentrated under vacuum. The final crude is separated by column chromatography (SiO$_2$, hexanes/DCM 100:0 → 70:30), affording the product as a yellowish powder (395 mg, 83% yield). **$^1$H NMR (400 MHz, CD$_2$Cl$_2$)**: δ = 7.53 (m, 8H, ArC-**H**), 7.12 (d, 2H, $^3J$ = 16.6 Hz, H-C=C-**H**), 7.07 (d, 2H, $^3J$ = 16.6 Hz, **H**-C=C-H), 7.08 (s, 2H, **α H**) 3.23 (s, 2H, C≡C-**H**) ppm; **$^{13}$C NMR (101 MHz, CD$_2$Cl$_2$)**: δ = 137.84, 137.57, 137.55, 137.42, 137.03, 137.02, 136.91, 135.57, 135.34, 135.28, 134.49, 134.39, 133.84, 133.81, 133.78, 132.91, 132.80, 132.60, 127.23, 124.70, 122.78, 83.63 (C≡C-H), 78.78 (C≡C-H), 57.25 (**α C**) ppm; **ATR-IR:** ν (cm$^{-1}$) = 3297 (w) (C≡C-H), 3073 (w) (ArC-H), 3035 (w) (ArC-H), 2109 (w) (C≡C), 1635 (w) (C=C), 1603 (w), 1541 (w) (ArC-ArC), 1506 (w) (ArC-ArC), 1465 (w), 1363 (m) (Cl-ArC-ArC-Cl), 1342 (m) (Cl-ArC-ArC-Cl), 1290 (s) (Cl-ArC-ArC-

Cl), 1262 (m), 1234 (w), 1212 (m), 1140 (m), 1107 (w), 1082 (w), 1016 (w), 967 (m), 947 (m), 910 (w), 859 (w), 836 (w), 817 (s) (C-Cl), 807 (s), 767 (w), 726 (w), 707 (w), 694 (w), 672 (w), 655 (m), 642 (w), 621 (w), 612 (w); **LDI-ToF (negative mode):** $m/z$ 942.6 [M – H], 870.7 [M – 2 Cl – H], 834.7 [M – 3 Cl – H]; **UV/Vis (THF):** λ (nm) (ε) 236 (57241), 255 (36647), 316 (62467)

### *4.3. Synthesis of compound 1-Rad*

All process was carried out in dark. Compound **1-H** (102 mg, 0.120 mmol, 1 eq) is dissolved in THF (4.5 mL) previously filtered through neutral $Al_2O_3$. 54-56 % aqueous $Bu_4NOH$ (100 μl, 0.234 mmol, 1.95 eq) is added and the solution is stirred. The formation of the perchlorotriphenylmethyl anion is monitored by UV/vis spectroscopy. When the deprotonation is complete, *p*-chloranil is added (63 mg, 0.256 mmol, 2.13 eq) and the oxidation from the perchlorotriphenylmethyl anion to the radical is followed by UV/vis spectroscopy. When the oxidation is complete, the mixture is evaporated under vacuum and the crude is purified by column chromatography ($SiO_2$, hexanes/$CH_2Cl_2$ 9:1) to afford the pure compound as a dark green powder (76 mg, 75% yield). **ATR-IR:** ν ($cm^{-1}$) = 3300 (w) (C≡C-H), 3081 (w) (ArC-H), 3035 (w) (ArC-H), 2105 (w) (C≡C), 1627 (w) (C=C), 1600 (w), 1507 (w) (ArC-ArC), 1462 (w) (ArC-ArC), 1413 (w), 1380 (s), 1361 (s), 1336 (m) (Cl-ArC-ArC-Cl), 1320 (m) (Cl-ArC-ArC-Cl), 1293 (m) (Cl-ArC-ArC-Cl), 1258 (m) (Cl-ArC-ArC-Cl), 1225 (m) (Cl-ArC-ArC-Cl), 1157 (w), 1120 (w), 1107 (w), 1083 (w), 1047 (w), 1029 (w), 1016 (w), 967 (m), 945 (m), 908 (w), 875 (w), 859 (w), 815 (s) (C-Cl), 768 (w), 735 (m), 709 (m), 693 (m), 666 (m), 652 (s), 642 (m), 606 (m), 577 (w); **LDI-ToF (negative mode):** $m/z$ 850.8 [M], 780.9 [M – 2 Cl]; **UV/Vis (THF):** λ (nm) (ε) 238 (44476), 305 (24461), 369 (18238), 385 (27801), 418 (14088), 437 (14246), 573 (1593); **CV (0.1 M TBAPF$_6$ in THF, vs Ag/AgCl):** $E_{1/2}$ = 0.02 V; **EPR:** g = 2.0026, a($^1H_a$) = 1.9 G, $\Delta H_{pp}$ = 1.1 G, a($^{13}C_{Ar}$) = 12.6, 14.3 G, a($^{13}C_{orto}$) = 14.3 G, a($^{13}C_\alpha$) = 29.4 G, a($^{13}C_{para}$) = 12.6 G; **SQUID:** C = 0.381 $cm^3$ K $mol^{-1}$, g = 2.0161.

### *4.4. Synthesis of compound 2-Rad*

All process was carried out in dark. Compound **2-H** (152 mg, 0.161 mmol, 1 eq) is dissolved in THF (6 ml) previously filtered through neutral $Al_2O_3$. 54-56 % aqueous $Bu_4NOH$ (95 μl, 0.2 mmol, 1.2 eq) is added and the solution is stirred. The formation of the perchlorotriphenylmethyl anion is monitored by UV/vis spectroscopy. When the deprotonation is complete, *p*-chloranil is added (50 mg, 0.2 mmol, 1.2 eq) and the oxidation from the perchlorotriphenylmethyl anion to the radical is followed by UV/vis spectroscopy. When the oxidation is complete, the mixture is evaporated under vacuum and the crude is purified by column chromatography (SiO$_2$, hexanes/ CH$_2$Cl$_2$ 100:0 → 70:30) to afford the pure compound as a dark green powder (106 mg, 70% yield). **ATR-IR:** ν (cm$^{-1}$) = 3296 (m) (C≡C-H), 3074 (w) (ArC-H), 3034 (w) (ArC-H), 2107 (w) (C≡C), 1626 (w) (C=C), 1559 (w), 1506 (m) (ArC-ArC), 1464 (w) (ArC-ArC), 1413 (w), 1381 (w), 1338 (s) (Cl-ArC-ArC-Cl), 1320 (s) (Cl-ArC-ArC-Cl), 1292 (m) (Cl-ArC-ArC-Cl), 1258 (m) (Cl-ArC-ArC-Cl), 1212 (m) (Cl-ArC-ArC-Cl), 1175 (w), 1162 (m), 1150 (w), 1125 (w), 1110 (w), 1082 (w), 1052 (w), 1016 (w), 964 (m), 946 (m), 921 (w), 859 (w), 838 (w), 816 (s) (C-Cl), 772 (w), 735 (m), 727 (w), 708 (m), 697 (w), 657 (m), 642 (w), 610 (w), 573 (w), 536 (w), 518 (w); **LDI-ToF (negative mode):** *m/z* 942.6 [M], 870.7 [M – 2 Cl]; **UV/Vis (THF):** λ (nm) (ε) 237 (48601), 256 (35726), 306 (48831), 378 (20206), 407 (21321), 441 (27435), 592 (15287) ; **CV (0.1 M TBAPF$_6$ in THF, vs Fc$^+$/Fc):** E$_{1/2}$ = -0.57 V; **EPR:** g = 2.0026, a($^1$H$_a$) = 1.9 G, a($^1$H$_b$) = 0.4 G, ΔH$_{pp}$ = 1.0 G, a($^{13}$C$_{Ar}$) = 12.6, 15.0 G, a($^{13}$C$_\alpha$) = 29.3 G, a($^{13}$C$_{para}$) = 12.6 G. **SQUID:** C = 0.378 cm$^3$ K mol$^{-1}$, g = 2.0069.

## 5. Experimental details

### 5.1. Conductance Measurements in Mechanically-Controlled Break Junctions

Mechanically-Controlled Break Junctions (MCBJ) are made on a phosphorus bronze flexible substrate coated with a polyimide insulating layer. A gold nanowire with a constriction is defined on top by electron beam lithography. The polyimide underneath the gold constriction is removed by reactive ion etching resulting in suspended gold bridge. Atomically sharp electrodes with nanometer-scale separations are formed by bending the flexible substrate in a three points bending mechanism. The conductance (G = I/V) through the wire is recorded during this process as a function of the electrodes displacement; when the Au-Au point contact is broken a molecule can bridge the two remaining electrodes, forming an Au-molecule-Au junction, the conductance then shows a molecule-dependent plateau behaviour with conductance below 1 G$_0$ (2e$^2$/*h*), the quantum of conductance. Thousands of traces are analysed and we present the information in the form of logarithm-binned two-dimensional (2D) and one-dimensional (1D) histograms. The former one maps how often a particular value of conductance is measured at a particular electrode displacement; the second one integrates the over the displacement, giving information of the most probable conductance values. A 2 μl drop of **2-Rad** molecules in dichloromethane (50 μM) is drop-casted on a clean gold nanowire and then pumped down to $10^{-5}$ mbar.

*5.2. Scanning Tunnel Microscopy (STM) break junction*

The conductance measurements of PTM-bisthiophene single-molecule junctions were carried out using the MCBJ described before and electrochemical Scanning Tunneling Microscopy-Break Junction (STM-BJ) technique. The experimental details and instrumentation descriptions were given in our previous publications.[5],[6] Briefly, the STM-BJ experiments were performed by a modified Molecular Imaging PicoSPM. The commercial STM scanner was modified with a dual channel preamplifier.[5] The current signal was converted to two voltage signals with the conversion factors ~21 μA/V and 10 nA/V. Both signals were split to the original as well as the 10 times amplified signal. The custom-designed program drives the STM tip at a controlled rate (87 nm/s) toward the STM substrate when the feedback is switched off. The approaching tip was stopped for a few ms when the upper current limit was reached (e.g. 10 $G_0$). The tip is then withdrawn at a controlled rate from the substrate to ensure the complete breaking of the molecular junction. These cycles were repeated hundreds of times. The current-time traces of each cycle (retraction part) were recorded with the digital oscilloscope (Yokogawa DL750). The raw data were analyzed with a lab-made program WA-BJ implemented in LabView2011. All the linear channels were combined to an integrated trace after manipulation by a respective gain factor. Finally, all the histograms were generated without any data selection.

The gold STM tips were prepared by electrochemical etching of gold wires (99.999%, 0.25 mm diameter) in a 1:1 (v/v) mixture of 30% HCl and ethanol.

*5.3. 1-Rad SAM preparation and characterization*

The SAMs were prepared on gold substrates evaporated on mica with a gold thickness of 300 nm. The substrates were first rinsed in acetone, dichloromethane and ethanol, placed in an ozone chamber for 20 minutes and then immersed in ethanol for further 30 minutes. Finally they are rinsed with isopropanol and dried under a nitrogen stream. The freshly cleaned substrates were immersed in 0.65 mM solution of the corresponding compound in dry toluene, in the dark, under argon atmosphere and mild heating (40 °C) the first 6 h, and at room temperature the following 42 h to maximize surface coverage. Then, the substrates were rinsed with copious amounts of toluene to ensure the complete elimination of any physisorbed material, and dried under nitrogen stream.

5.3.1. Cyclic voltammetry (CV)

The surface coverage was obtained from the conventional equation: $\Gamma = Q/nFA$ where $\Gamma$ is the surface concentrations in mol cm$^{-2}$, Q is the charge consumed in the reduction or oxidation process and it is the integrated area of the anodic or cathodic charge, n is the number of electrons transferred (n=1), F is Faraday's constant, and A is the electrode surface area.

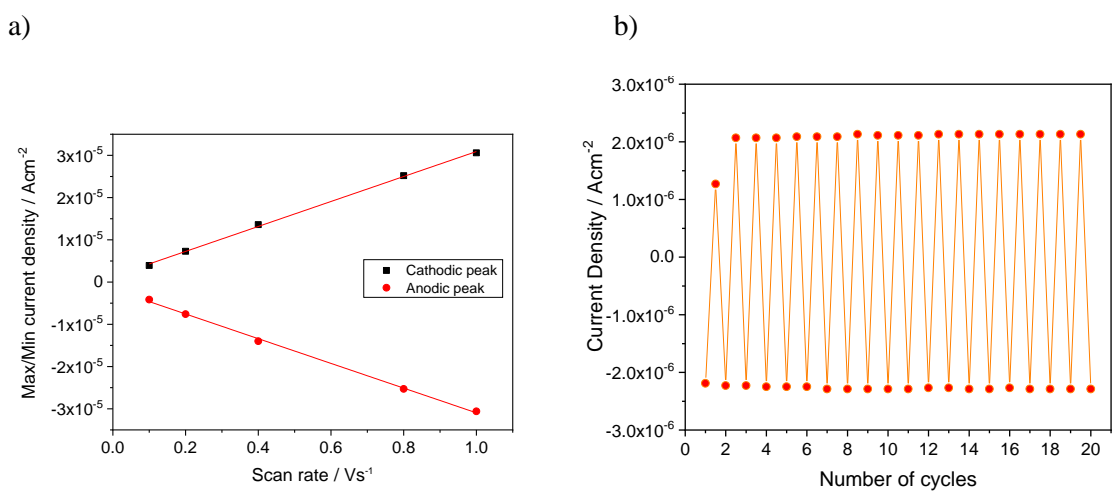

**Figure S2.** a) **P**lot of the peak current density *versus* scan rate for the cathodic and anodic peak from Figure 2c in the main text b) maximum current densities upon 20 sweeping voltage cycles at 0.1 V/s. Working electrode: **SAM-1-Rad**, Counter electrode: Pt wire, Reference electrode: Ag wire, Internal reference: Ferrocene, Electrolyte: 0.1 M tetrabutylammonium hexafluorophosphate in THF.

*5.3.2. X-ray photoelectron spectroscopy (XPS)*

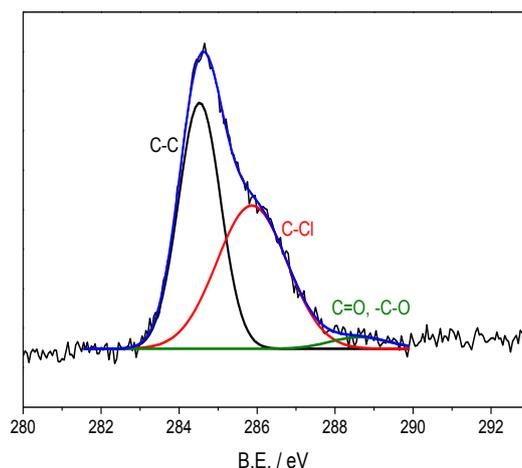

**Figure S3.** Deconvoluted XPS spectrum of C1s of **SAM-1-Rad**.

The C1s shows a peak at 284.5 eV assigned to C-C bond, a peak at higher-energy, 285.9 eV, attributed to C-Cl and a very low intense peak at 288.6 eV that could be due to some oxidized carbon.

*5.3.3 Transport through the SAMs.*

The liquid metal GaOx/EGaIn was used as the top-electrode since it is known that it forms a non-invasive soft top-contact with the SAM. For these studies SAMs of **1-Rad** (Scheme S1, page S6) were prepared on ultra-smooth template-stripped gold (Au$^{TS}$) bottom electrodes and followed by the formation of the top-electrodes following previously reported methods.[Ref. 7 and references therein] The top-electrode is biased and the bottom electrode is grounded. We collected at least 12 traces on 36 different junctions for three samples (i.e. on three **1-Rad** SAMs on Au). In total, 432 traces J(V) were recorded. Exemplary J(V) (1 trace = 0V→1.4V→-1.4V→0) curves for three different samples (to check the reproducibility) are shown in Figure S4a. In order to obtain representative values from the J(V)s the data were treated statistically to determine the mean, the standard deviation and the 95% confidence intervals. For each potential value (voltage steps of 0.1 V) the data were plotted in histograms in $\log_{10}|J|$ and a Gaussian fit was applied. The results of these statistical treatments have been used to plot the Figure S4b, where the points represent the mean value and the error bar represents the 95% confidence interval as a function of the applied potential.

Compound **1-H** (Scheme S1, page S6) that is the non-radical precursor of **1-Rad** has also been employed to prepare SAMs on Au$^{TS}$. These SAMs were used for comparison with the **1-Rad** SAM and to compare it with our previous reported results. The SAM preparation methodology explained above for **1-Rad** was followed for **1-H**. As observed in Figure S4b the measured tunneling current through the **1-H** SAM is lower, here one order of magnitude, if compared to the radical based monolayer. This is in accordance with our previous work [Ref.7] where we postulated that the charge transport mechanism implies the participation of the single unoccupied molecular orbital (SUMO) of the grafted radical that lowers the tunneling barrier height.

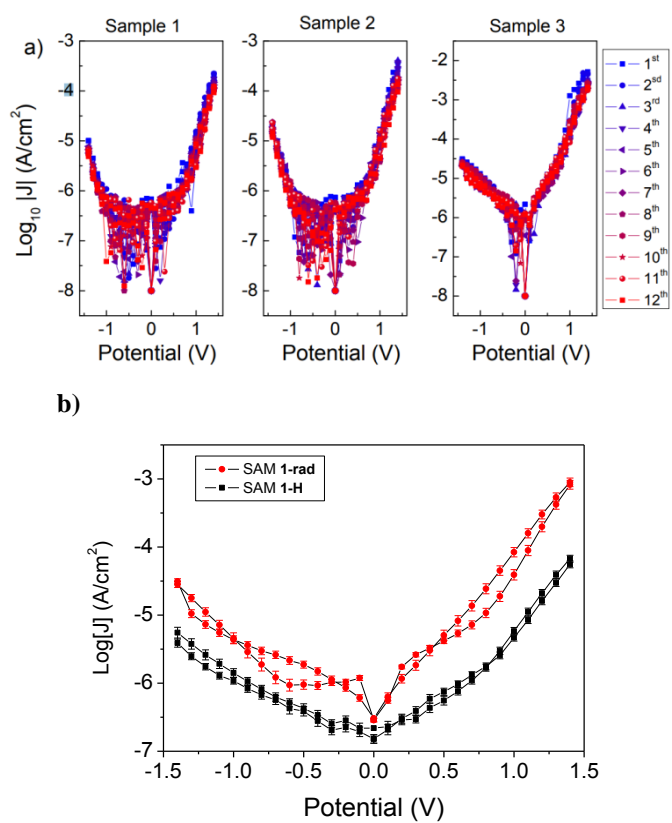

**Figure S4**. a) Exemplary J(V) measurements on a given junction for three different samples (i.e. three different **1-Rad** SAMs). b) Mean values and their corresponding 95% confidence interval as a function of voltage for **1-Rad** and **1-H** SAMs.

*5.4. Additional electron transport characterization of 2-Rad molecular junctions*

Figure S5 shows the characteristic length distribution obtained for the **2-Rad** molecular junction reported in the main text. The large narrow peak at 0.5 nm is due to direct tunneling between the electrodes as explained in the previous section. A second maximum in the length distribution appears centered at 2.3 nm. The total stretching of the Au/**2-Rad**/Au molecular junction, taking into account the snap-back correction $z^* = 0.5$ nm, is therefore $z = 2.8$ nm. This value approximately matches the length of the stretched **2-Rad** molecule, about 2.7 nm.

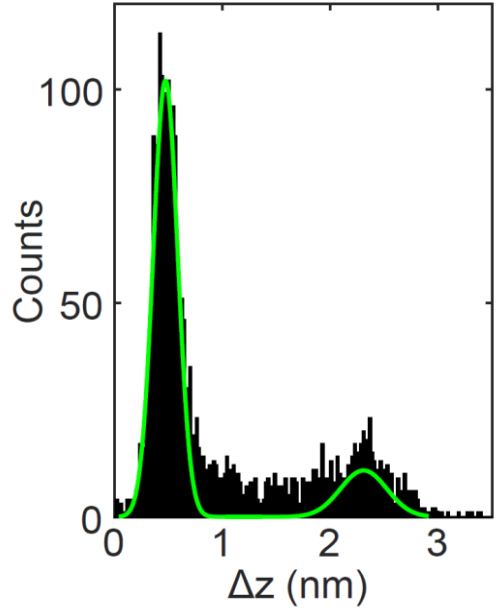

**Figure S5.** Characteristic length distribution for a Au/**2-Rad**/Au molecular junctions

*5.5 Low-temperature electron-transport measurements in Au/2-Rad/Au MCBJ: Kondo effect*

Figure S6 shows some selected dI/dV curves measured at T = 4 K in a mechanically-controlled break junction setup. The current is measured at fixed inter-electrode distances and the differential conductance is thereafter numerically obtained. A zero-bias resonance is observed in around a 10% of the junctions. This resonance is typically associated to Kondo correlations in molecules or quantum dots containing magnetic impurities and appears when the electronic hybridization of impurity and conduction electrons is strong. The small sharp dip centered at around V = 0 V is an artifact caused at low bias voltages by the use of a logarithmic amplifier.

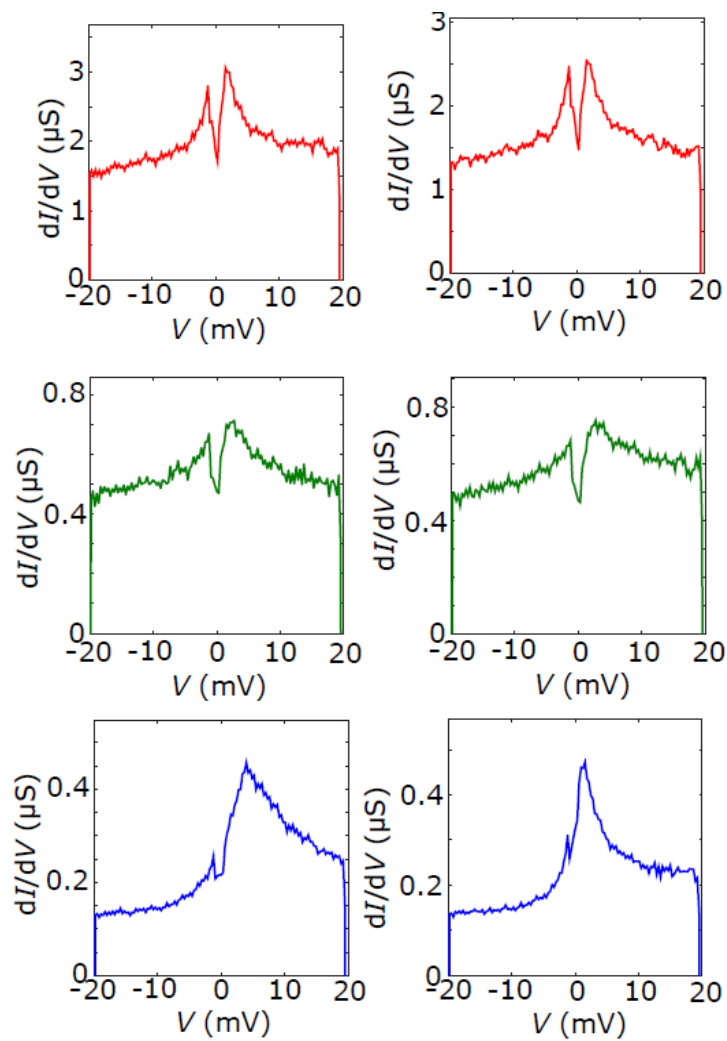

**Figure S6**. Differential conductance (d*I*/d*V*) versus bias voltage *V* curves measured in three junctions at two fixed inter-electrode distances and at *T* = 4 K. A zero-bias resonance, fingerprint of Kondo correlations, is observed. The small sharp dip centered at V = 0 V is an artifact of the measurement due to the use of a logarithmic amplifier at low-bias voltages.

Note that the intensity of the Kondo effect in transport, at a fixed temperature, depends exponentially on: (a) the charging energy *U*, (b) the molecule-electrode electronic coupling *Γ* and (c) the molecule's level alignment to the electrodes' Fermi energy *ε*. It is well known that a slight variation in any of these parameters, or the combination of them, drastically modifies the Kondo temperature and therefore the observability of the Kondo peak. Note also that the lower statistical conductance of the **2-Rad** with respect to the **3-Rad** could be indicative of a lower *Γ* or a larger *ε*. This may be translated in average lower Kondo temperatures that in turn would difficulty the observation of Kondo effect at 4 K, the lower limit of typical MCBJ setups. More statistics and further analysis in temperature and magnetic field would be required to shed light on this aspect.

S15

## 5.6. Details and electron transport measurements on radical PTM-bisthiophene molecule

### 5.6.1. Chemical structure of the PTM-bisthiophene molecule

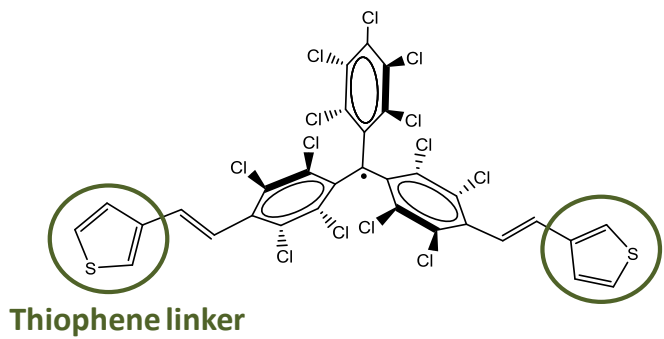

**Figure S7.** Chemical structure of the radical **PTM-bisthiophene** molecule (**3Rad** in the main text).

### 5.6.2. Electron transport measurements at room temperature by STM-BJ

Conductance of **PTM-bisthiophene** was measured at room temperature by STM-BJ technique. Two conductance peaks $G_L$ and $G_H$ (*Low* and *High*) were resolved. Fig. S8a shows some representative individual traces. The trace *1* displays only the contribution from electron tunnelling through a medium (in this case argon) and no molecular plateau. The traces *2* and *3* have a molecular plateau around $10^{-4}$ $G_0$ ($G_H$), *4* and *5* display both plateaus at $G_H$ and $G_L$, and *6* and *7* have only $G_L$. The plateau at $G_L$ and $G_H$ could appear alone or together. The maximum of length distribution (Figs. S8c and S8d) for $G_H$ feature is at $\Delta z^* = 0.45$ nm. The peak centered at 0.12 nm in Fig. S8d is related to electron tunnelling through the medium (argon). The junction probability for $G_H$ in drop-cast experiment was around 50% as estimated from the area of the peak fits in Fig. S8d.

Using the value $\Delta z_{corr} = 0.50$ nm as estimated for snap-back of gold electrodes after breaking gold-gold contact, we can calculate the average value of the actual stretching distance of Au/ **PTM-bisthiophene**/Au junctions according to $z = \Delta z^* + z_{corr}$ [8]. For **PTM-bisthiophene**, we obtain $z \sim 0.95$ nm. This is significantly smaller than the length of the molecule, which is ~1.7 nm (sulphur to sulphur distance of the PTM-bisthiophene molecule). These results are rationalized by means of DFT calculations. See Section 5.7

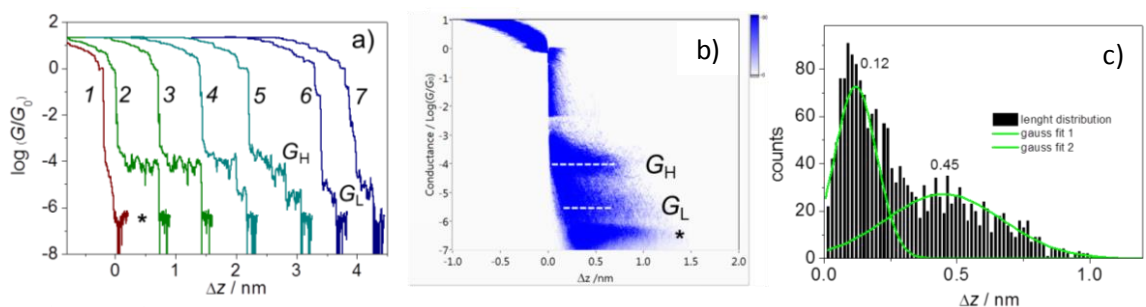

**Figure S8.** (a) Individual conductive traces without and with molecular plateaus for **PTM-bisthiophene**, obtained by STM-BJ with $V_{bias}$ = 0.1 V and a stretching rate of 87 nm s$^{-1}$. (b) 2D conductance histograms for **PTM-bisthiophene**. (c) Characteristic length distribution between $10^{-4.6}$ G$_0$ and $10^{-0.3}$ G$_0$. The measurements were done at room temperature in argon atmosphere after drop casting 20 μL of 0.08 mM **PTM**-**bisthiophene** in dichloromethane and drying in a gentle argon stream.

## 6. DFT calculations

### 6. DFT calculations

### 6.1. Computational details

DFT was used to estimate the adsorption energy of the thiophene linkers on Au, the rapture force of the C-Au covalent bond, and to optimize the geometry of the Au/molecule/Au junction for the transport simulations. DFT calculations were performed in a supercell approach by using the all-electron code FHI-AIMS [9]. The standard numerical atom-centered orbitals basis set "tier 1" was considered for the Au atoms, while all atoms in the molecules (Cl, C, S, H) were described by using the "tier 2" basis sets. Negligible basis superposition error was found when computing adsorption energies and potential energy curves. This is in accordance with the tests reported in Ref. [9] and the results of previous works [10]. The generalized gradient approximation (GGA) by Perdew, Burke and Ernzerhof (PBE) [11] was adopted for the exchange-correlation density functional. Van der Waals interactions (vdW) were included by combining the PBE functional with the screened vdW interaction method for molecules on surfaces (called vdW$^{surf}$) [12,13]. The used vdW parameters for Au are the same as in Ref. 14. Energy differences were converged with respect to the number of *k*-points within 1 meV.

*6.2. Adsorption energy of the thiophene linker on gold.*

The adsorption of a thiophene linker on Au was modelled by placing a single thiophene molecule on the top fcc surface of a 5-layers Au slab with a (3x3) square unit cell. About 70 Å of vacuum separated the slab from its periodic image. The adsorption energy is defined as $E_{ad} = E_{mol+Au} - (E_{mol} + E_{Au})$, where $E_{mol+Au}$ is the energy of the combined molecule-slab system, while $E_{mol}$ and $E_{Au}$ are respectively the energy of the isolated molecule and slab contained in the same supercell. The adsorption energy was obtained as function of the molecule-surface distance $r$. The results are reported in Fig. S9 for the molecule in the vertical-laying configuration (V-CFG) and in Fig. S10 for the molecule in the horizontal-laying configuration (H-CFG). In both cases the thiophene S atom is on top of an Au surface atom (i.e. top adsorption site). By using only GGA we found that, at the equilibrium adsorption distance, $E_{ad}$ = -0.06 eV and $E_{ad}$ = -0.31 for V-CFG and H-CFG respectively, while GGA+vdW$^{surf}$ gives $E_{ad}$ = -0.15 eV and $E_{ad}$ = -0.74 eV, respectively. In both configurations, the molecule is therefore mostly physisorbed with a large fraction of the adsorption energy (about 85 %) that is due to the van der Waals contribution. We remind that GGA-PBE may spuriously predict a small binding even for purely physisorbed systems because of the approximate exchange energy [15] (in contrast the exact exchange energy will give no binding at all as vdW interactions arise from correlation effects [16]).

The result of GGA+vdW$^{surf}$ for the adsorption energy of H-CFG at the equilibrium adsorption distance is very close to the range of experimental values at low coverage (from -0.57 eV to -0.68 eV) [17, 18] and is in striking agreement with the result obtained by hybrid DFT/Møller–Plesset Perturbation Theory calculations (-0.73 eV) [19]. In contrast, as shown in other works [20,21], semi-empirical corrections for the vdW energy, such as the DFT-D approach, return much larger (in absolute value) adsorption energies, which are far off the experimental range [19].

Finally, we point out that, when H-CFG is fully relaxed until the atomic forces become smaller than 0.01 eV/ Å, a small angle (4.5$^O$) forms between the molecule and the surface plane, with the thiophene S-atom pointing toward the surface. For such geometry, the adsorption energy is only slightly increased (in absolute value) to -0.77 eV.

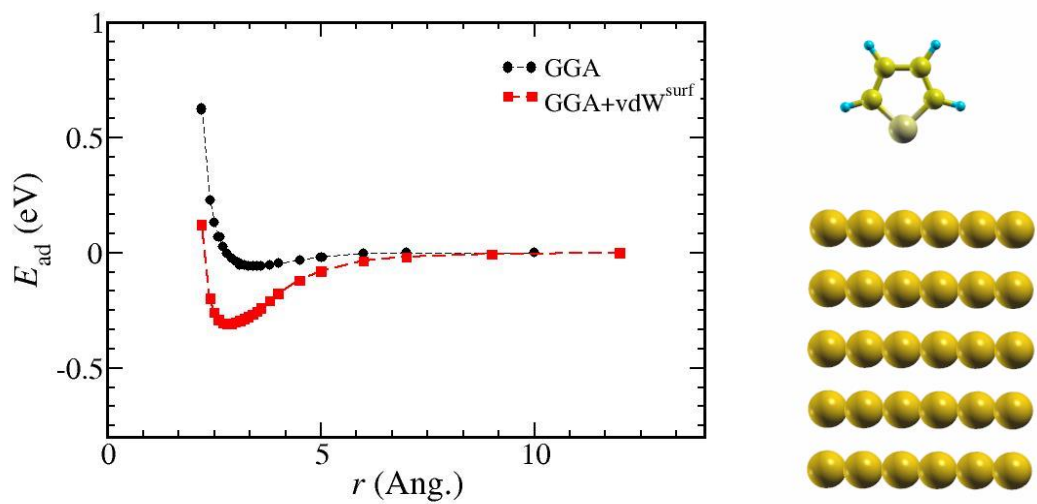

**Figure S9.** Left: Adsorption energy for a thiophene molecule on gold in the vertical configuration as function of the molecule (S atom)-surface distance. Right: The slab-molecule system used in the calculations.

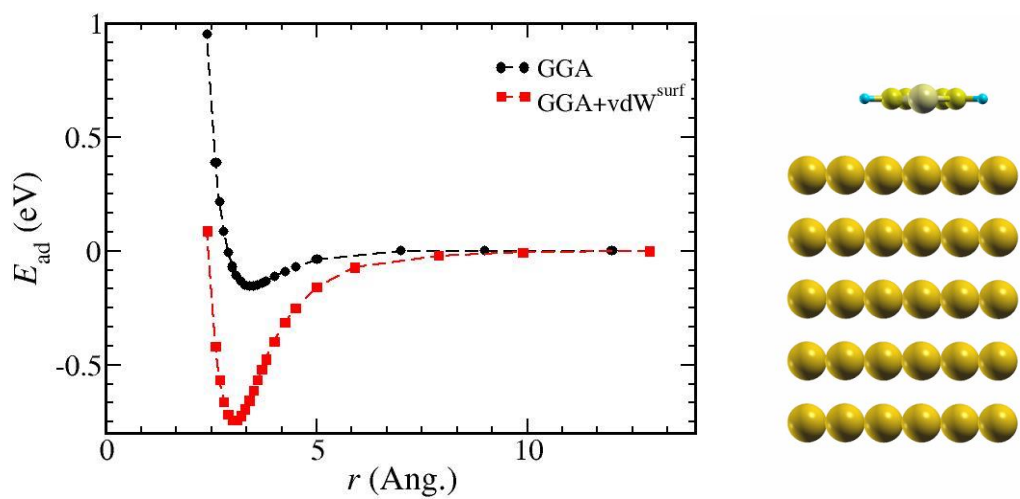

**Figure S10.** Left: Adsorption energy for a thiophene molecule on gold in the horizontal configuration as function of the molecule-surface distance. Right: The slab-molecule system used in the calculations.

### *6.3. Rupture force*

In order to estimate the rupture force of the C-Au and of the S-Au bond, we considered single phenylacetylene, phenylthiol and thiophene molecules attached to the surface of the 5-layers Au slab that was previously used for the adsorption energy calculations (see Fig. S11, S12, S13 – right panels). In case of phenylacetylene and phenylthiol the H atom of the acetylene and of the thiol group was removed, so that the unsaturated C and S atoms could form a covalent bond with the surface as in typical devices (Fig.1 main text). In case of thiophene, since the covalent bond with a flat Au surface is extremely weak as explained in Sec. 6.2, we considered the bond of the thiophene S atom with an Au surface adatom, which was placed on the hollow site of the Au surface and whose then position was optimized.

We first calculated the potential energy surface (PES) by elongating the C(S)-Au bond by discrete steps $\Delta r$ of either 0.05 or 0.1 Å around the equilibrium bond-length $r_e$, while keeping the relative positions of the atoms in the molecule as well as in the slab constrained (Fig. S11, S12, S13). Note that the calculations were performed without including vdW interactions, since we are interest in the strength of the covalent bonds. Afterwards we then fitted the PES by means of a Morse potential

$$V_{Morse}(r) = D\left(1 - e^{-2Fr/D}\right)^2,$$

where $r = r_e + \Delta r$, $D = V(\infty) - V(r_e)$ defines the bond dissociation energy and $F$ the maximum rupture force. Finally, following Ref. 22 we computed the effective potential associated to the bond when it is stretched from the equilibrium bond length $r_e$ to a larger length $r_s$ because of an external force $F_s = V'(r_s)$. That effective potential reads

$$V_{eff}(r) = V(r) - rV'(r_s).$$

The local maximum of $V_{eff}$ at $r = r_{TS}$ separating the attractive from the repulsive part is the transition state of the bond breaking event. The activation energy for the bond breaking is then given by $E_a = V_{eff}(r_{TS}) - V_{eff}(r_s)$ (neglecting the contribution from the zero point energy). By increasing $r_s$, the attractive part of $V_{eff}$ tend to vanish, and when $r_s$ approaches the corresponding $r_{TS}$, the activation energy becomes comparable to or smaller than room temperature (Fig. S11).

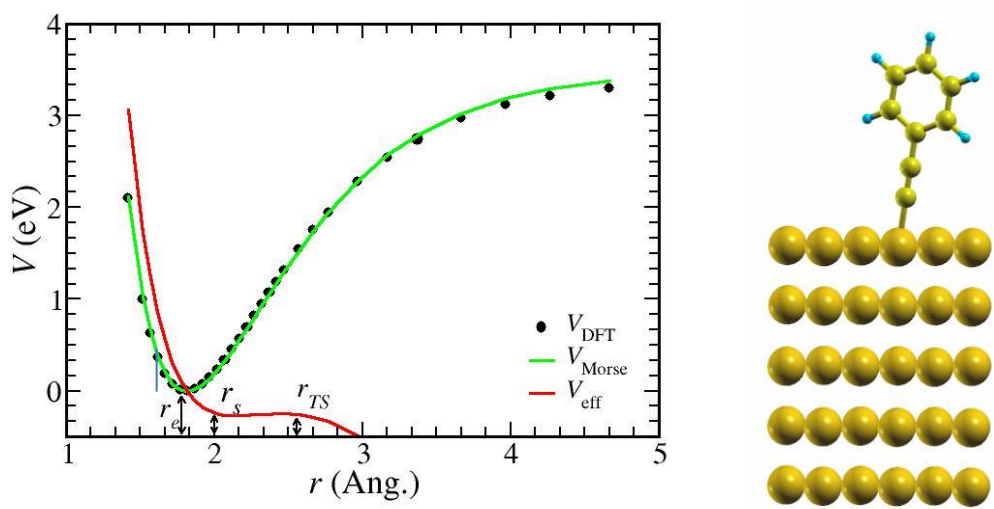

**Figure S11.** Right: Phenylacetylene adsorbed on the gold slab. Left: DFT potential energy surface (black open dots), fitted Morse potential (green line) and effective potential for $r_s = 0.3$ Å when the activation energy is equal to room temperature.

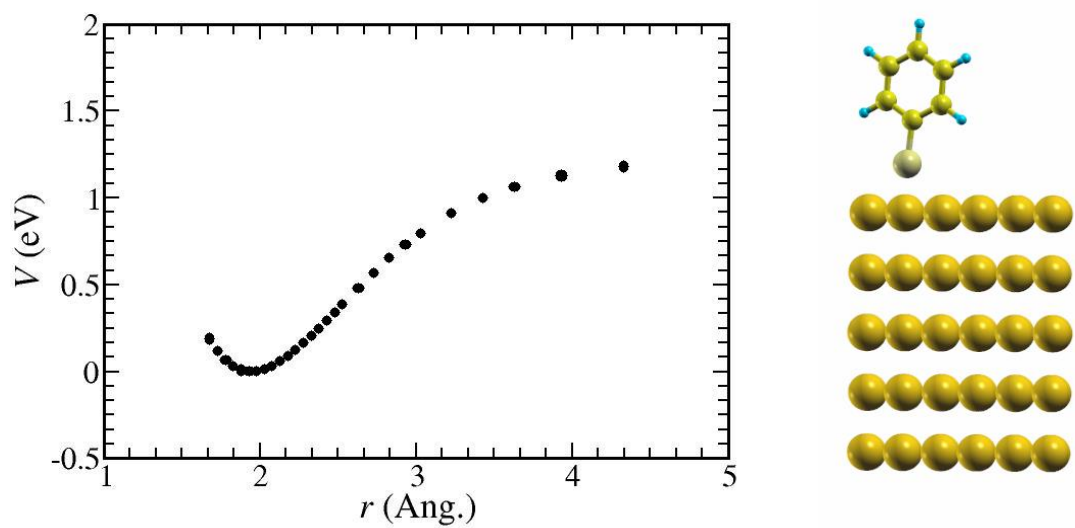

**Figure S12.** Right: Phenylthiol adsorbed on the gold slab. Left: DFT potential energy surface.

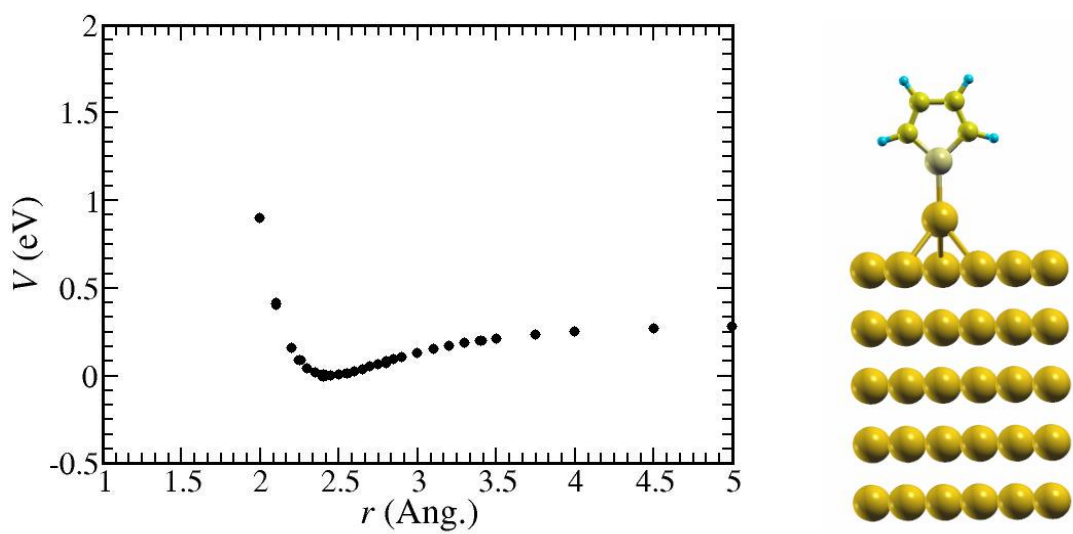

**Figure S13.** Right: Thiophene adsorbed on the gold slab with an adatom. Left: DFT potential energy surface.

## 7. Conductance Calculations

The conductance is calculated by using the *Smeagol* first principles electron transport code [23, 24], which combines density functional theory (DFT) with the non-equilibrium Green's function (NEGF) approach. The DFT Hamiltonian is obtained from the Siesta code [25]. We use the local density approximation (LDA) for the exchange correlation potential, and apply a scissor operator to correct the energy gap for the frontier molecular orbitals, with details specified in Ref. 26. The real-space mesh is set by an equivalent energy cutoff of 300 Ry, and we use a standard double-ζ plus polarization basis set for all atoms. We verified that due to the large size of the supercell in the plane we can evaluate the electronic structure at the Γ-point in the Brillouin zone perpendicular to the transport direction.

The atomic structures for the electron transport calculations are obtained from structural relaxations of the central part of the scattering region performed by using the DFT. The computational details are explained in Sec 6.1 and are analogous to those used in previous works [27]. The molecule and all the Au atoms, except for those at the boundary that are directly connected to leads in the transport setup, are allowed to relax until forces are smaller than 0.01 eV/Å.

# ANEX I: ¹H-NMR, ¹³C-NMR, FT-IR, UV-Vis, LDI-TOF, EPR and CV spectra of the synthesized molecules

(*E*)-6,6'-((2,3,5,6-tetrachloro-4-(4-ethynylstyryl)phenyl)methylene)bis(1,2,3,4,5-pentachlorobenzene) (**1-H**)

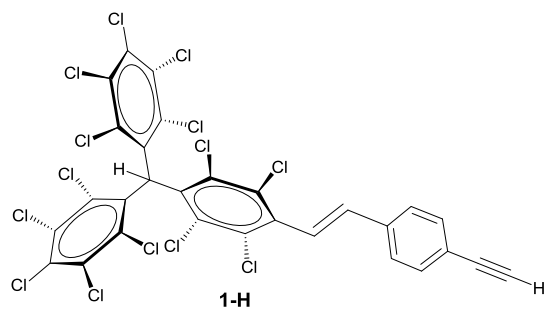

¹H NMR, CD$_2$Cl$_2$, 400 MHz

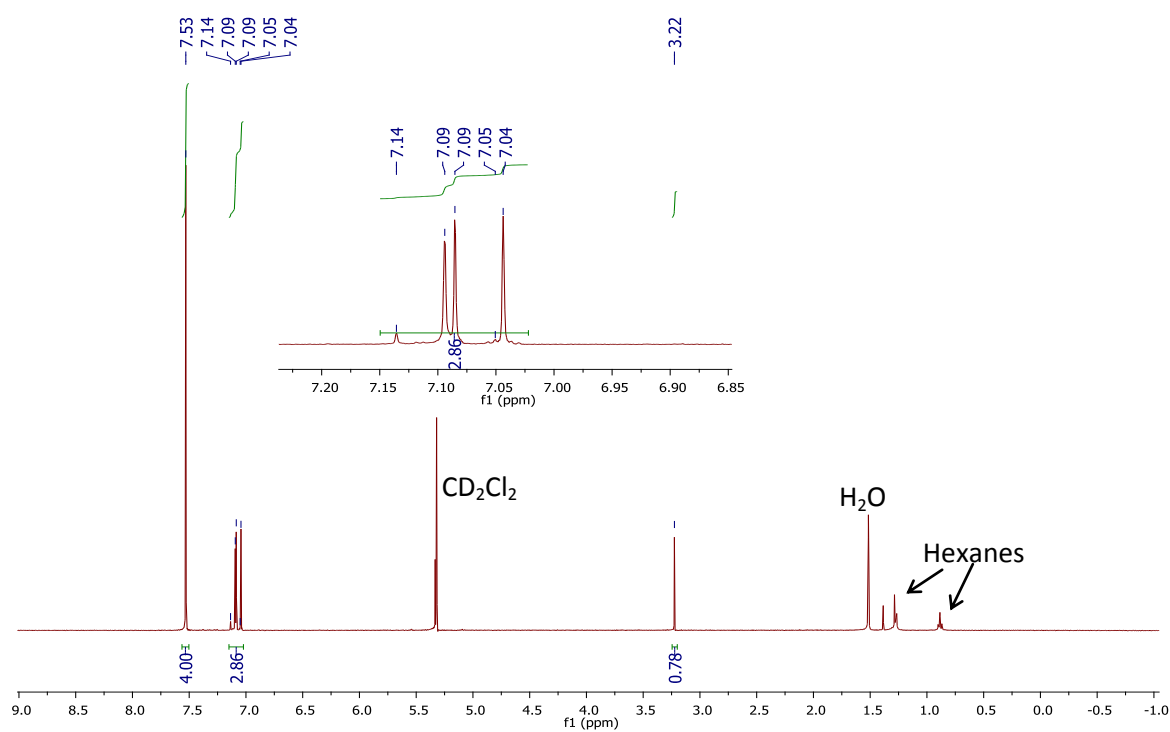

**Figure S14.** ¹H NMR spectrum of compound **1-H** in CD$_2$Cl$_2$.

¹³C NMR, CDCl₃, 101 MHz

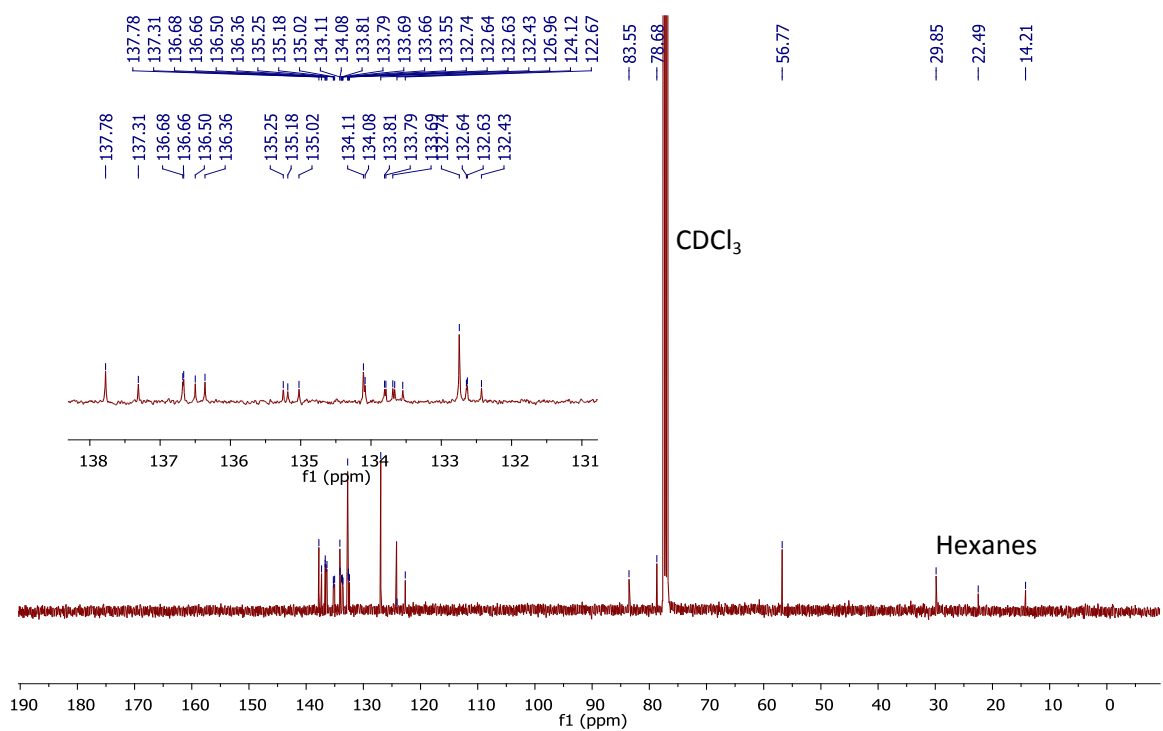

**Figure S15.** ¹H-NMR spectrum of compound **1-H** in CDCl₃.

FT-IR

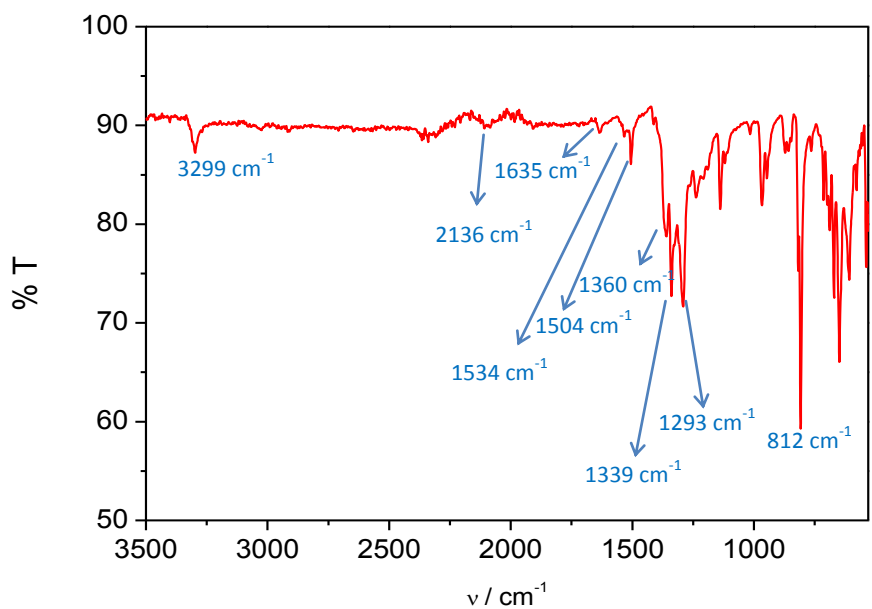

**Figure S16.** FT-IR spectrum of compound **1-H** in powder.

UV/vis

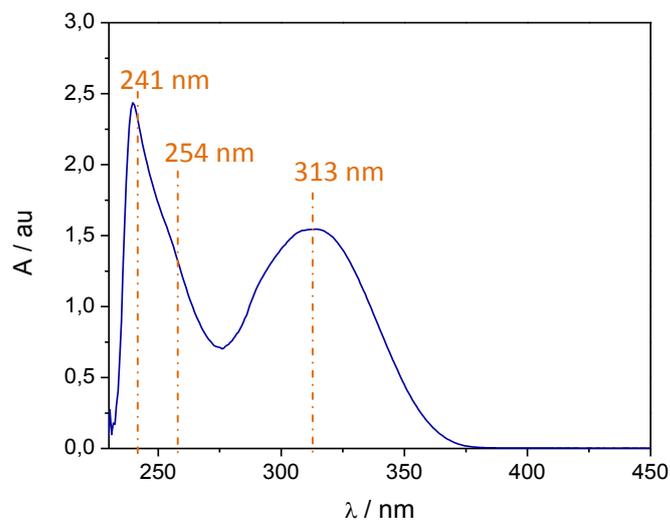

**Figure S17.** UV/vis spectrum of compound **1-H** in THF (4.734·10$^{-5}$ M)

LDI-ToF

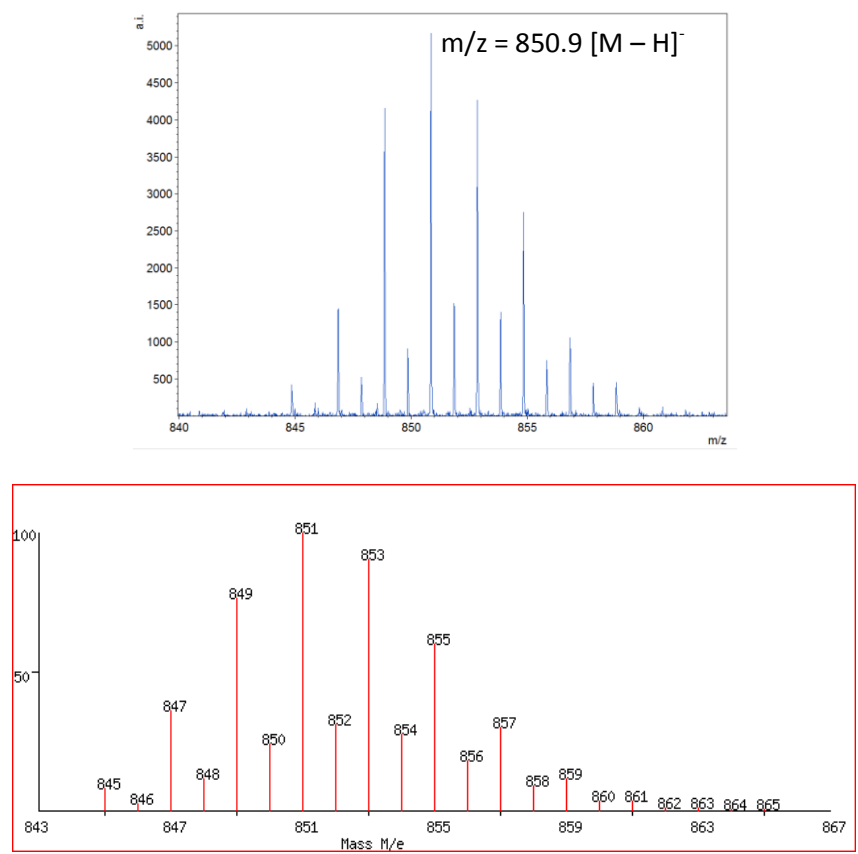

**Figure S18.** LDI-ToF spectra of compound **1-H** experimental (top) and simulated (bottom).

**6,6'-((perchlorophenyl)methylene)bis(1,2,4,5-tetrachloro-3-((*E*)-4-ethynylstyryl)benzene) (2-H)**

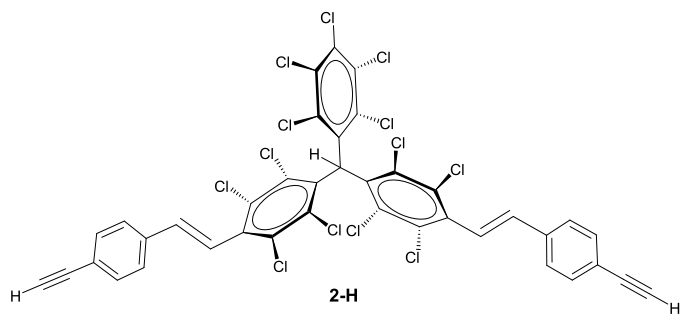

¹H NMR, CD$_2$Cl$_2$, 400 MHz

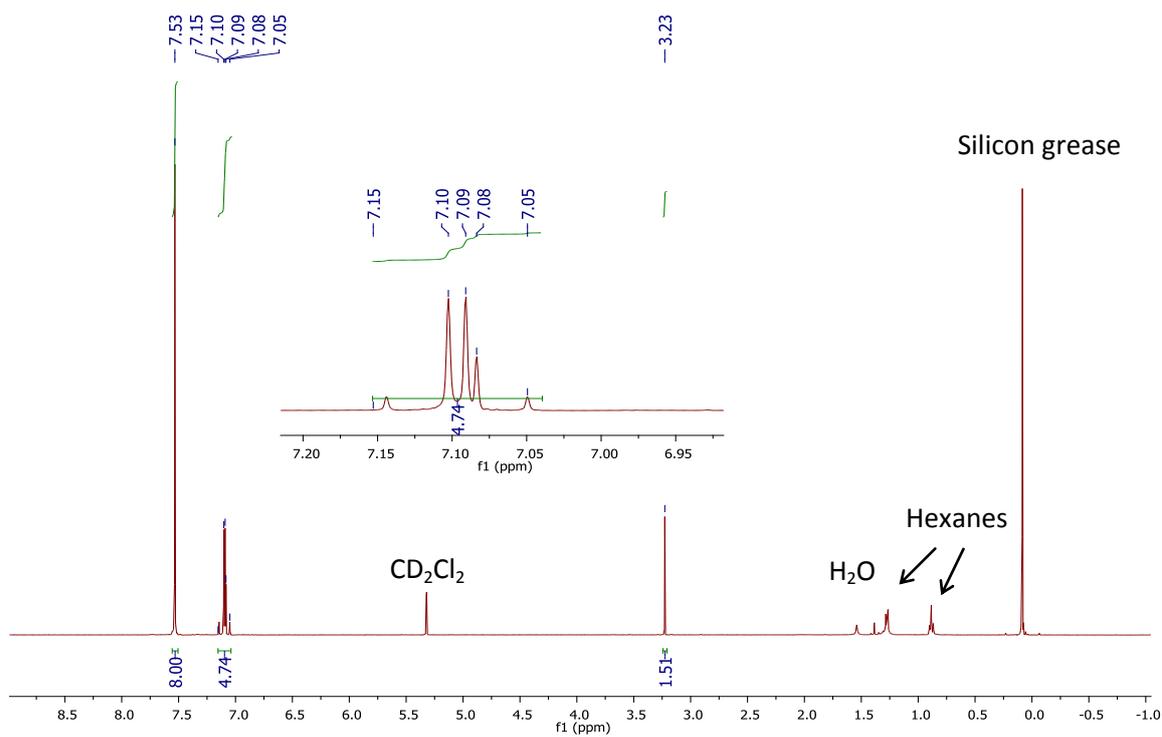

**Figure S19.** ¹H NMR spectrum of compound **2-H** in CD$_2$Cl$_2$.

¹³C-NMR, CD₂Cl₂, 101 MHz

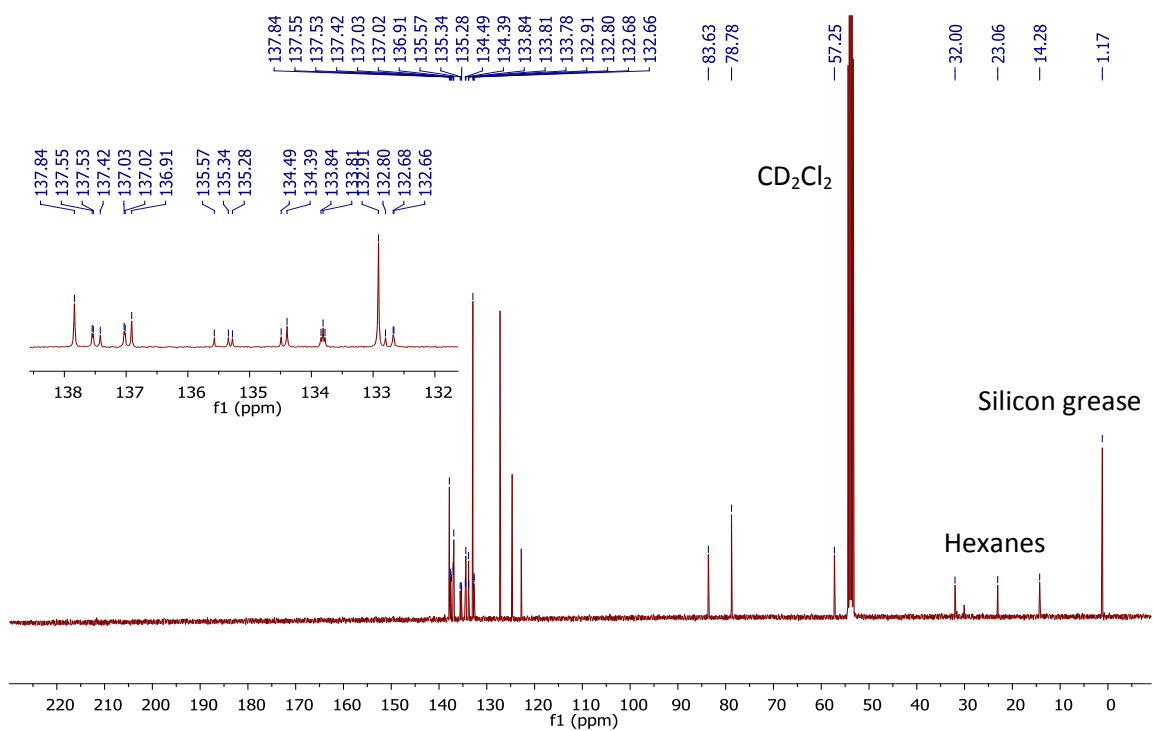

**Figure S20.** ¹³C NMR spectrum of compound **2-H** in CD₂Cl₂.

FT-IR

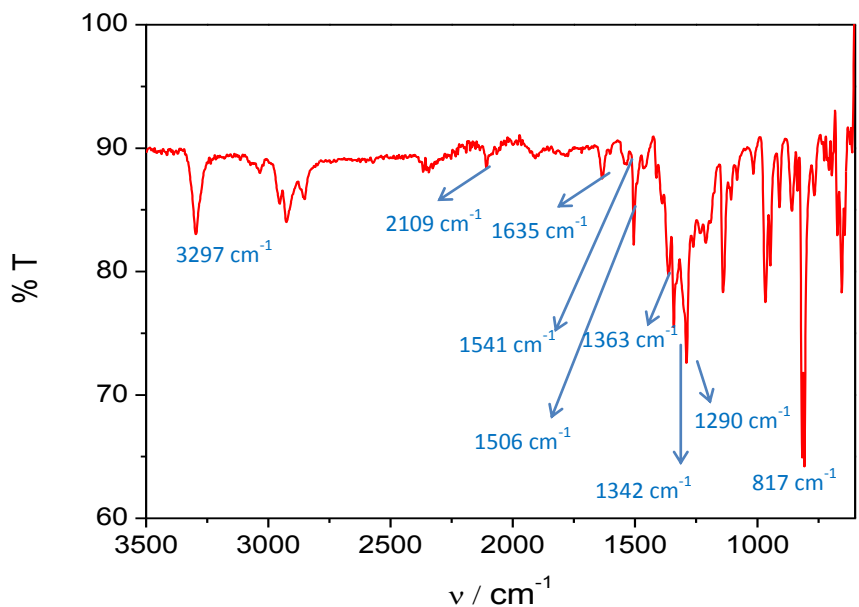

**Figure S21.** FT-IR spectrum of compound **2-H** in powder.

UV/vis

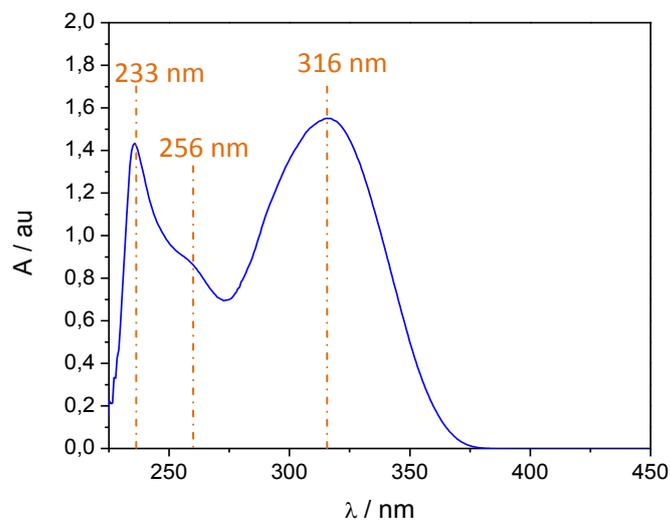

**Figure S22.** UV/vis spectrum of compound **2-H** in THF (2.50·10$^{-5}$ M)

LDI-ToF

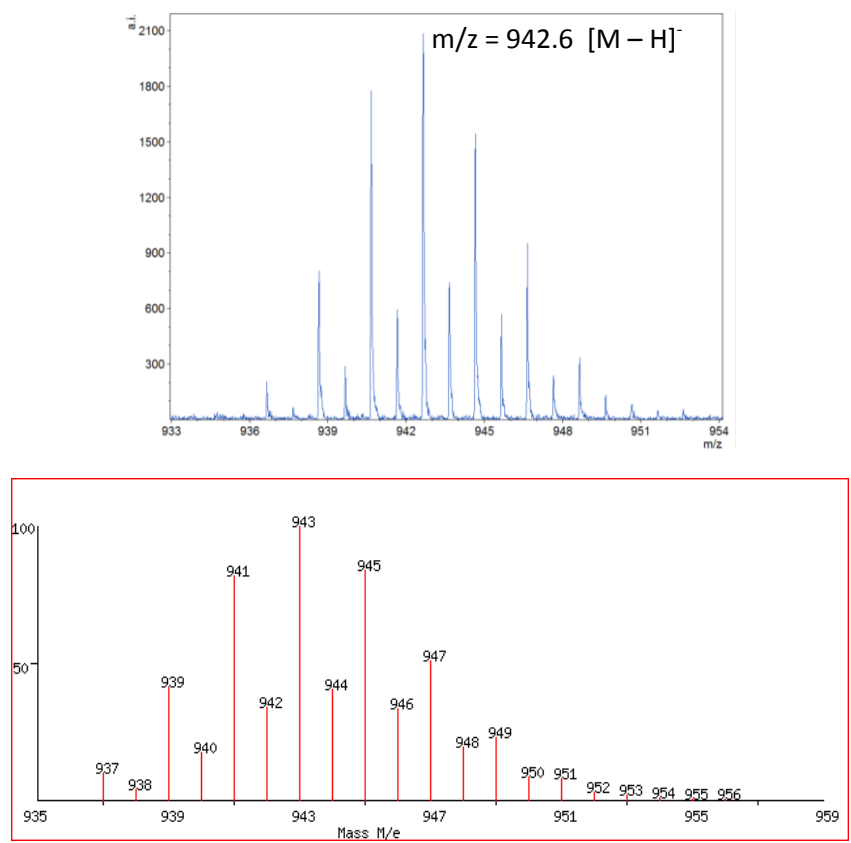

**Figure S23.** LDI-ToF spectra of compound **2-H** experimental (top) and simulated (bottom).

**(*E*)-6,6'-((2,3,5,6-tetrachloro-4-(4-ethynylstyryl)phenyl)methylene)bis(1,2,3,4,5-pentachlorobenzene) radical (1-Rad)**

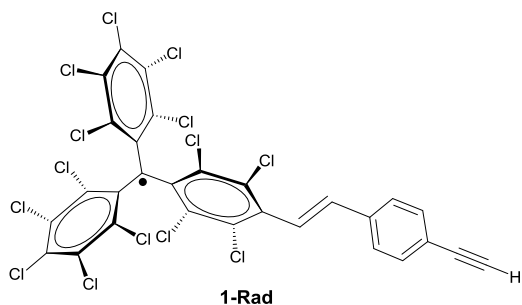

FT-IR

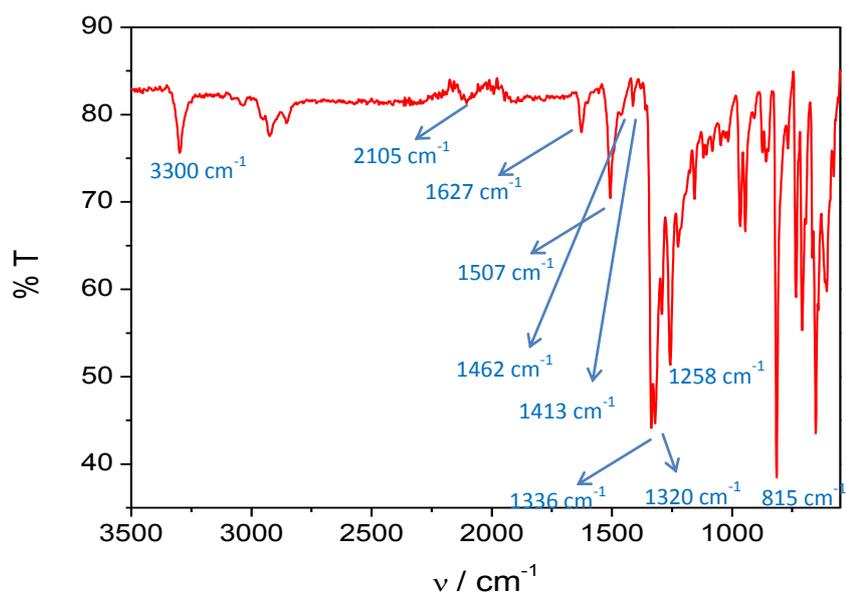

**Figure S24.** FT-IR spectrum of compound **1-Rad** in powder.

UV/Vis

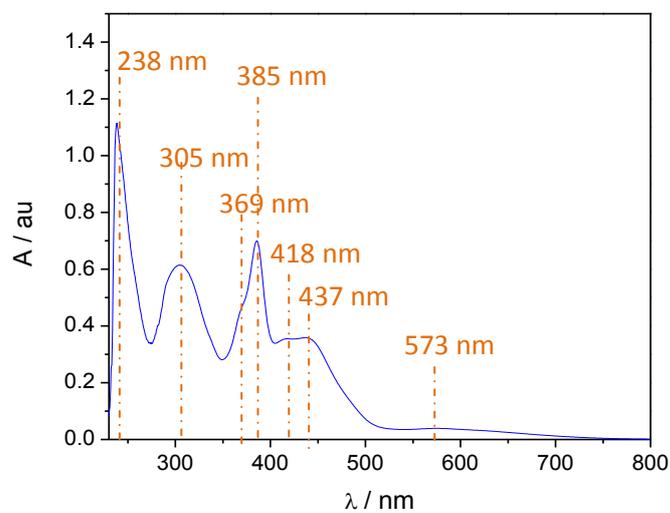

**Figure S25.** UV/vis spectrum of compound **1-Rad** in THF (2.52·10$^{-5}$ M)

EPR

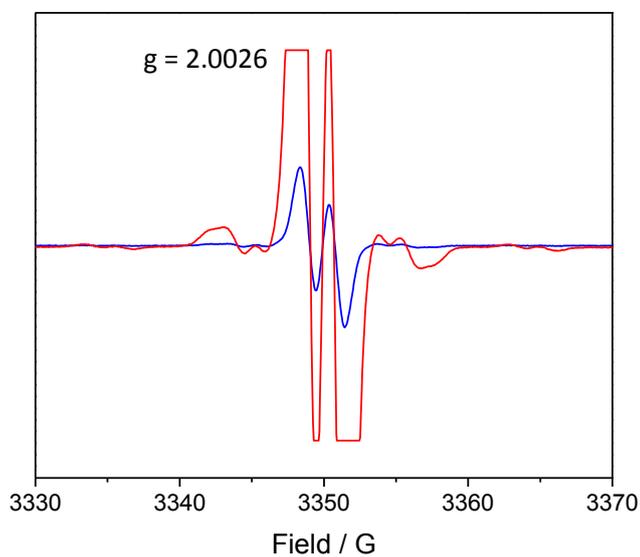

**Figure S26.** EPR spectrum of compound **1-Rad** in dichloromethane under normal (blue) and strong (red) recording conditions.

SQUID

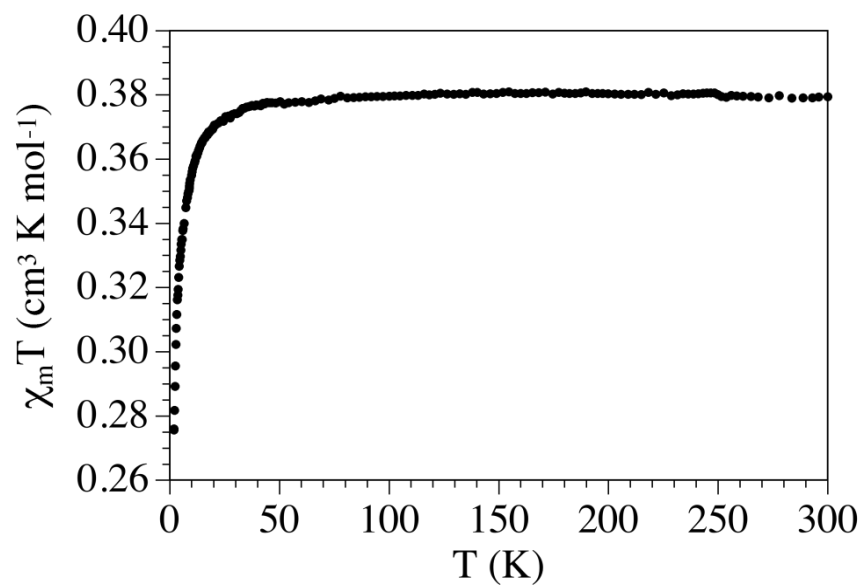

**Figure S27.** Temperature dependence of the magnetic susceptibility of **1-Rad**.

CV

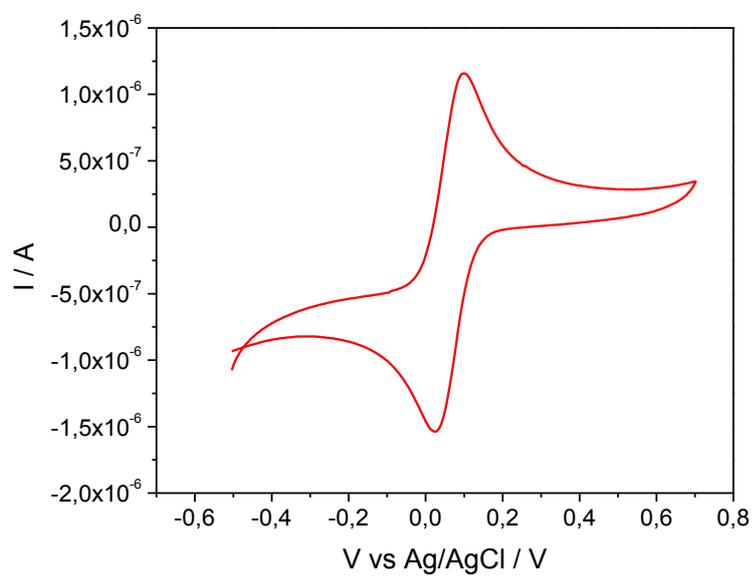

**Figure S28.** Cyclic voltammetry of compound **1-Rad** in a 0.1 M TBAPF$_6$ in THF solution. Electrodes: Pt wire as working, Pt wire as counter and Ag/AgCl as reference.

LDI-ToF

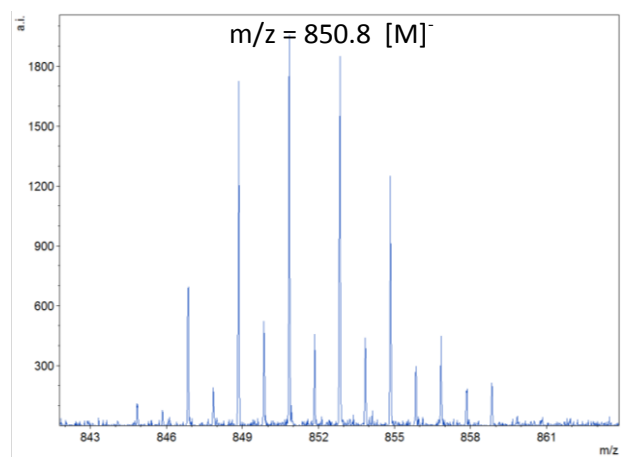

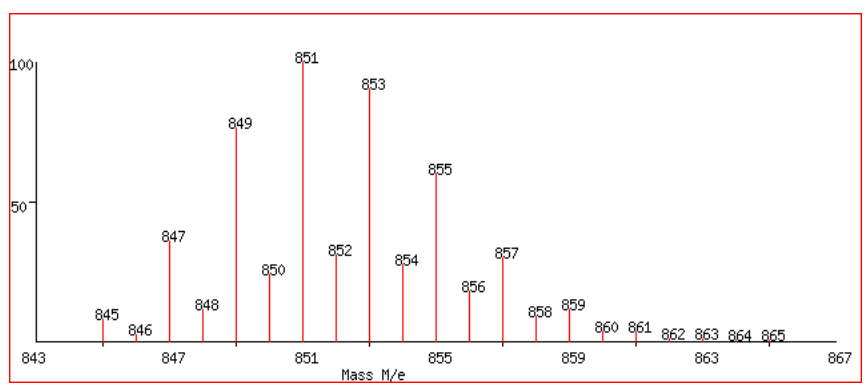

**Figure S29.** LDI-ToF spectra of compound **1-Rad** experimental (top) and simulated (bottom).

**6,6'-((perchlorophenyl)methylene)bis(1,2,4,5-tetrachloro-3-((*E*)-4-ethynylstyryl)benzene) radical (2-Rad)**

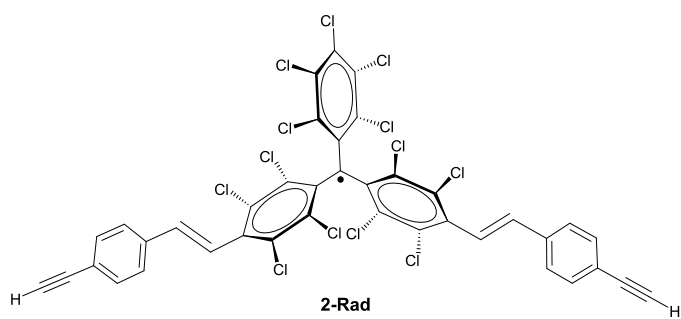

FT-IR

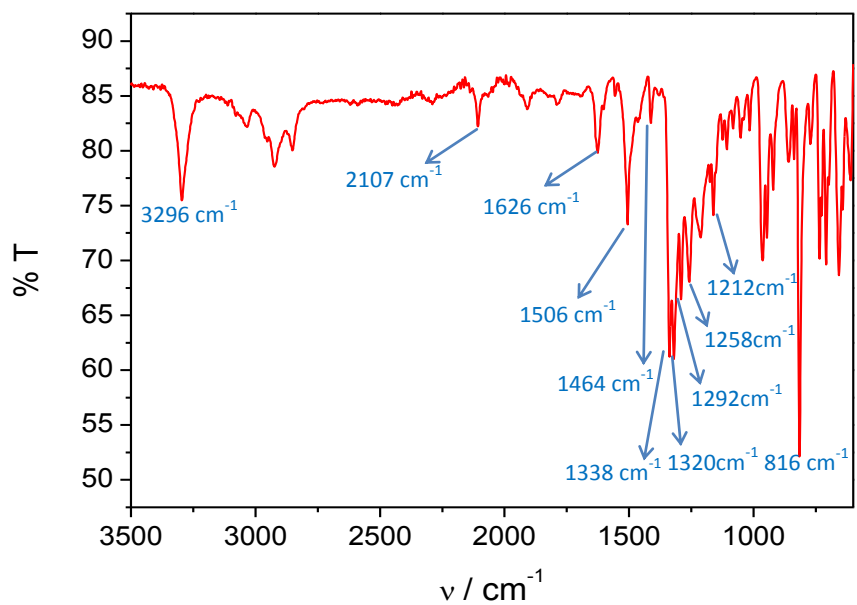

**Figure S30.** FT-IR spectrum of compound **2-Rad** in powder.

UV/vis

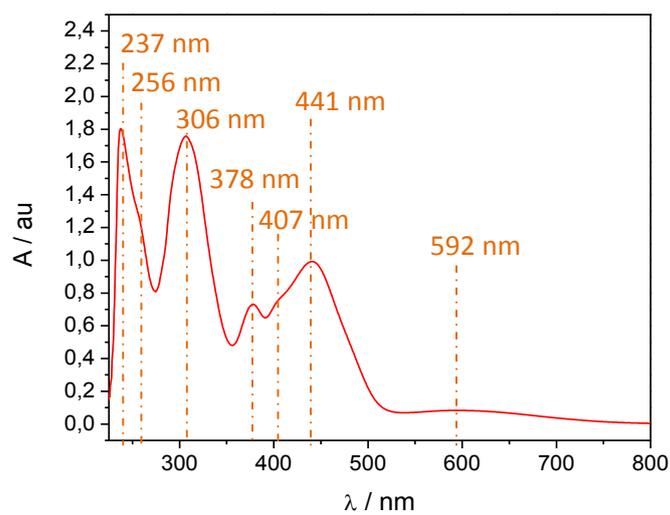

**Figure S31.** UV/vis spectrum of compound **2-Rad** in THF ($3.64 \cdot 10^{-5}$ M).

EPR

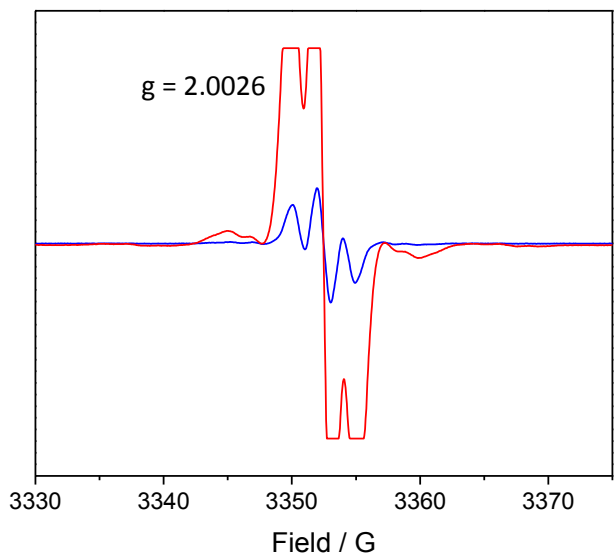

**Figure S32.** EPR spectra of compound **2-Rad** in dichloromethane under normal (blue) and strong (red) recording conditions.

SQUID

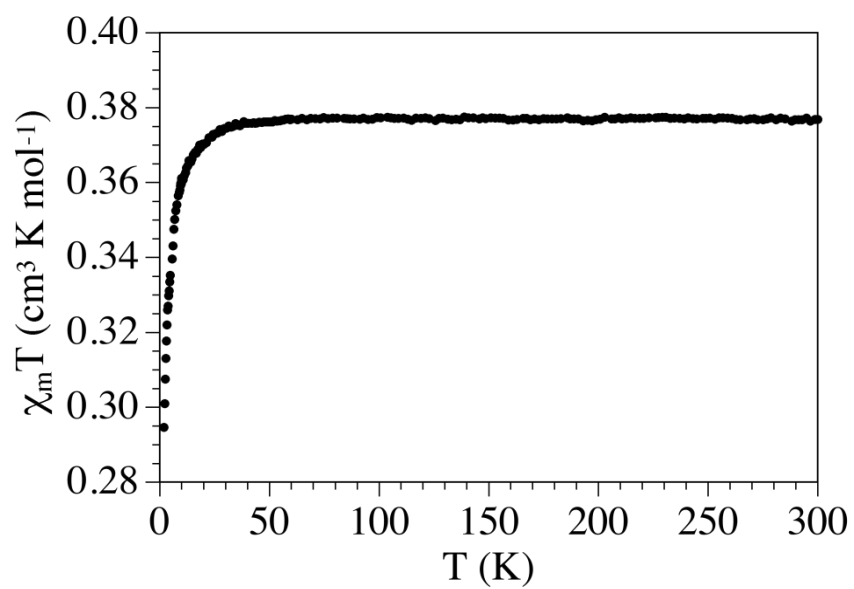

**Figure S33.** Temperature dependence of the magnetic susceptibility of **2-Rad**.

CV

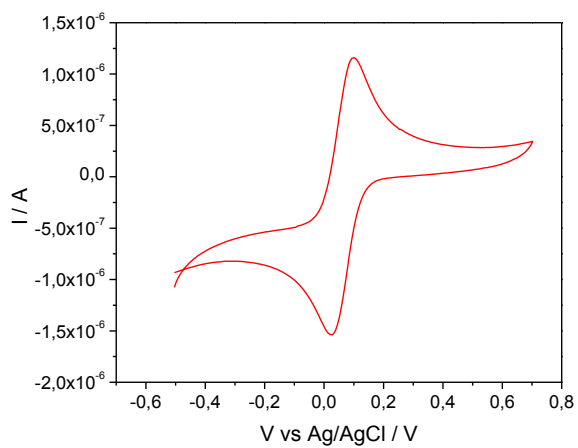

**Figure S34.** Cyclic voltammetry of compound **2-Rad** in a 0.1 M TBAPF$_6$ in THF solution. Electrodes: Pt wire as working, Pt wire as counter and Ag/AgCl as reference.

LDI-ToF

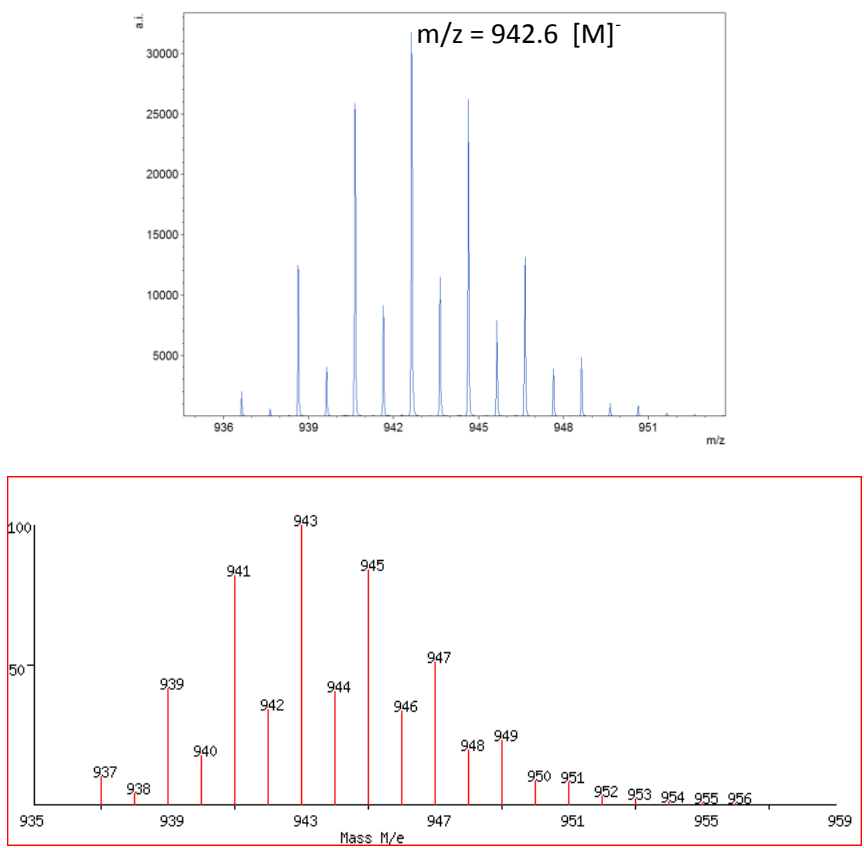

**Figure S35.** LDI-ToF spectra of compound **2-Rad** experimental (top) and simulated (bottom).